\documentclass[prl, twocolumn, floatfix, nofootinbib]{revtex4-2}
\usepackage{graphicx}
\usepackage{amsmath,amssymb,tikz,physics,mathtools,bm}
\usepackage[colorlinks=true, allcolors=purple]{hyperref}

\usetikzlibrary{decorations.pathreplacing}

\makeatletter 
    
\renewcommand\onecolumngrid{
\do@columngrid{one}{\@ne}%
\def\set@footnotewidth{\onecolumngrid}
\def\footnoterule{\kern-6pt\hrule width 1.5in\kern6pt}%
}

\renewcommand\twocolumngrid{
        \def\footnoterule{
        \dimen@\skip\footins\divide\dimen@\thr@@
        \kern-\dimen@\hrule width.5in\kern\dimen@}
        \do@columngrid{mlt}{\tw@}
}%

\makeatother

\newcommand{\bbra}[1]{\langle \! \langle #1 |}
\newcommand{\kett}[1]{| #1 \rangle \! \rangle}
\newcommand{\bbrakett}[2]{\langle \! \langle #1 | #2 \rangle \! \rangle}
\newcommand{\define}{\coloneqq}

\begin{document}

\title{Efficient Detection of Strong-To-Weak Spontaneous Symmetry Breaking via the R\'enyi-1 Correlator}

\author{Zack Weinstein}
\email{zackmweinstein@berkeley.edu}
\affiliation{Department of Physics, University of California, Berkeley, California 94720, USA}

\date{\today}

\begin{abstract}
Strong-to-weak spontaneous symmetry breaking (SWSSB) has recently emerged as a universal feature of quantum mixed-state phases of matter. While various information-theoretic diagnostics have been proposed to define and characterize SWSSB phases, relating these diagnostics to observables which can be efficiently and scalably probed on modern quantum devices remains challenging. Here we propose a new observable for SWSSB in mixed states, called the R\'enyi-1 correlator, which naturally suggests a route toward scalably detecting certain SWSSB phases in experiment. Specifically, if the canonical purification (CP) of a given mixed state can be reliably prepared, then SWSSB in the mixed state can be detected via ordinary two-point correlation functions in the CP state. We discuss several simple examples of CP states which can be efficiently prepared on quantum devices, and whose reduced density matrices exhibit SWSSB. The R\'enyi-1 correlator also satisfies several useful theoretical properties: it naturally inherits a stability theorem recently proven for the closely related fidelity correlator, and it directly defines SWSSB as a particular pattern of ordinary spontaneous symmetry breaking in the CP state. 
\end{abstract}

\maketitle


There has recently been an explosion of interest in characterizing and classifying mixed-state phases of quantum matter. Whereas the study of quantum phases has previously been largely restricted to quantum ground states or to thermal Gibbs states \cite{sachdevQuantumPhaseTransitions2011,Sachdev_2023}, experimental advances in quantum simulation \cite{altmanQuantumSimulatorsArchitectures2021,ebadiQuantumPhasesMatter2021,satzingerRealizingTopologicallyOrdered2021,semeghiniProbingTopologicalSpin2021,bluvsteinQuantumProcessorBased2022,leonardRealizationFractionalQuantum2023}, state preparation \cite{chenRealizingNishimoriTransition2023,iqbalTopologicalOrderMeasurements2024,iqbalNonAbelianTopologicalOrder2024,miStableQuantumcorrelatedManybody2024}, and error correction \cite{miaoOvercomingLeakageQuantum2023,acharya2024quantumerrorcorrectionsurface,bluvsteinLogicalQuantumProcessor2024} have motivated the study of a new class of ``locally decohered'' mixed states \cite{dennisTopologicalQuantumMemory2002,fan2024diagnostics,bao2023mixedstatetopologicalordererrorfield,lee2023quantumcriticality,zou2023channeling}. In turn, these states have revealed fresh theoretical insights on the relation between mixed-state topological order and quantum error correction \cite{chen2024separability,su2024tapestry,li2024replicatopologicalorderquantum,lu2024disentangling,li2024perturbativestabilityerrorcorrection,hauser2024informationdynamicsdecoheredquantum,sriram2024nonuniformnoiseratesgriffiths,sala2024stabilityloopmodelsdecohering,zhang2024strongtoweakspontaneousbreaking1form}, on new classifications of topological and symmetry-protected topological phases in mixed states \cite{lee2024symmetryprotectedtopologicalphases,ma2024topologicalphasesaveragesymmetries,wang2024intrinsicmixedstatequantumtopological,Guo2024twodimensional,sohal2024noisyapproachintrinsicallymixedstate,ellison2024classificationmixedstatetopologicalorders,zhang2024quantumcommunicationmixedstateorder}, and on the very notion of what constitutes a mixed-state phase of matter \cite{Coser2019classificationof,rakovszky2024definingstablephasesopen,sang2024mixed,sang2024stabilitymixedstatequantumphases,guo2024newframeworkquantumphases}. 

Traditionally, the most important characterization of both classical and quantum phases is how they manifest their symmetries \cite{landau1937theory,mcgreevyGeneralizedSymmetriesCondensed2023}. While a pure state can only be symmetric or nonsymmetric under a symmetry transformation, mixed states can be symmetric in one of two distinct senses: mixed states with a single well-defined symmetry charge are said to exhibit a ``strong'' symmetry, while mixed states composed of an incoherent mixture of symmetry charges exhibit only a ``weak'' symmetry. Curiously, there is a sense in which a mixed state's strong symmetry can be spontaneously broken \textit{without} breaking the corresponding weak symmetry \cite{lee2023quantumcriticality}. This phenomenon, recently dubbed \textit{strong-to-weak spontaneous symmetry breaking} (SWSSB) \cite{lee2023quantumcriticality,lessa2024strongtoweakspontaneoussymmetrybreaking,sala2024spontaneousstrongsymmetrybreaking,gu2024spontaneoussymmetrybreakingopen,huang2024hydrodynamicseffectivefieldtheory}, serves as a new universal characterization of phases of quantum matter which is unique to mixed states.

A crucial feature distinguishing SWSSB from conventional spontaneous symmetry breaking (SSB) is that its defining observables are necessarily information theoretic. 
%
%
In place of conventional ``linear'' observables, SWSSB has previously been defined by diagnostics such as the R\'enyi-2 correlator \cite{lee2023quantumcriticality,sala2024spontaneousstrongsymmetrybreaking} and the fidelity correlator \cite{lessa2024strongtoweakspontaneoussymmetrybreaking} (defined below), which measure the \textit{distinguishability} between a mixed state $\rho$ and the same state with added symmetry charges. While these diagnostics have several theoretically appealing features, they are also highly nonlinear in $\rho$ and extraordinarily difficult to measure experimentally. Consequently, the practical implications of SWSSB on observable features of mixed states in quantum devices have remained unclear.

In this Letter, we put forth a new observable for SWSSB called the \textit{R\'enyi-1 correlator}, denoted $R_1(x,y)$ below, which exhibits several theoretically and practically useful features. Whereas the R\'enyi-2 correlator is defined by treating $\rho$ as a pure state in a doubled Hilbert space, $R_1$ is defined in a doubled Hilbert space via two-point correlations in the \textit{canonical purification} (CP) of $\rho$. While $R_1$ exhibits much of the same useful symmetry properties of the R\'enyi-2 correlator, it also shares important information-theoretic features with the fidelity correlator. Specifically, $R_1$ directly inherits a \textit{stability theorem} recently proven for the fidelity correlator \cite{lessa2024strongtoweakspontaneoussymmetrybreaking}, which guarantees that a state with long-range order (LRO) in $R_1$ cannot be evolved to a state without such LRO by a strongly symmetric finite-depth channel.

Most importantly, the structural form of $R_1$ naturally suggests a simple method with which SWSSB can be directly and scalably observed in a large class of mixed-state experiments. If the CP of $\rho$ can be efficiently prepared, then measuring $R_1$ trivially amounts to measuring a two-point correlation function. If this correlator exhibits LRO while ordinary correlation functions of $\rho$ do not, then one can immediately conclude that $\rho$ exhibits SWSSB. Below we discuss several examples of CP states which can be prepared efficiently using standard techniques, and whose reduced density matrices exhibit SWSSB.

\textit{Strong-to-weak spontaneous symmetry breaking.---} A density matrix $\rho$ of a quantum system exhibits a strong symmetry under the group $G$ if there is a unitary representation $U_g$ of each $g \in G$ such that $U_g \rho = e^{i \varphi_g} \rho = \rho U_g$. In contrast, $\rho$ exhibits only a weak symmetry under $G$ if $U_g \rho U_g^{\dag} = \rho$ \cite{footnote_1}. For concreteness, we will primarily consider global on-site symmetries of many-body lattice spin models, where $U_g = \prod_j u_{j,g}$ factorizes across each local degree of freedom supported on the sites $j$ of a regular lattice, although generalizations to higher-form symmetries have also been considered \cite{zhang2024strongtoweakspontaneousbreaking1form,wang2024intrinsicmixedstatequantumtopological,sohal2024noisyapproachintrinsicallymixedstate,ellison2024classificationmixedstatetopologicalorders,kuno2024intrinsicmixedstatetopological}. Ordinary SSB in a state $\rho$ with (at least) weak $G$ symmetry is then defined as LRO in the two-point correlation function $\tr[ O_x O^{\dag}_y \rho ]$ of a local operator $O_x$ which is charged under the global symmetry, in the sense that $[O_x, U_g] \neq 0$ for some $g \in G$. In the mixed-state setting, if $\rho$ is strongly symmetric, it is said that such a state spontaneously breaks both the strong and weak symmetries.

Heuristically, a mixed state $\rho$ exhibits SWSSB if its strong symmetry is spontaneously broken without breaking the corresponding weak symmetry. To make this notion precise, several works have defined SWSSB using the ``R\'enyi-2'' correlator $R_2(x,y)$ \cite{lee2023quantumcriticality,bao2023mixedstatetopologicalordererrorfield,fan2024diagnostics,sala2024spontaneousstrongsymmetrybreaking}:
\begin{equation}
\label{eq:renyi_2}
		R_2(x,y) \coloneqq \frac{\tr[O_{xy} \rho O_{xy}^{\dag} \rho ]}{\tr \rho^2} = \frac{\bbra{\rho} O_{xy}^L \bar{O}_{xy}^R \kett{\rho} }{\bbrakett{\rho}{\rho}} ,
\end{equation}
where $O_{xy} \equiv O_x O^{\dag}_y$ and $\kett{\rho} \coloneqq (\rho \otimes 1) \sum_{\vb{s}} \ket{\vb{s}}^L \ket{\vb{s}}^R$ is the vectorization of $\rho$, with $\{\ket{\vb{s}}^{L,R}\}$ denoting complete product-state bases of identical ``left'' and ``right'' subsystems $L$ and $R$. The operators $O^L_{xy} \coloneqq O_{xy} \otimes 1$ and $\bar{O}^R_{xy} \coloneqq 1 \otimes O^*_{xy}$ respectively act only on the left and right subsystems. If $\rho$ exhibits a strong symmetry under $G$, the doubled state $\kett{\rho}$ exhibits a doubled symmetry under the group $G^L \times G^R$: for any $g, h \in G$, $U_g^L \bar{U}_h^R \kett{\rho} = e^{i(\varphi_g - \varphi_h)} | \rho \rangle \! \rangle$. Using this formalism, SWSSB has been previously defined as LRO in the R\'enyi-2 correlator \eqref{eq:renyi_2} in the \textit{absence} of LRO in the ordinary two-point correlation function $\tr [O_{xy} \rho]$ \cite{lee2023quantumcriticality,sala2024spontaneousstrongsymmetrybreaking}. In the doubled-state representation, the order parameter $O^L_x \bar{O}^R_x$ transforms nontrivially under the separate symmetries $G^L$ and $G^R$, but remains invariant \cite{footnote_2} under the ``diagonal'' subgroup $G^{\text{diag}}$ of $G^L \times G^R$ represented by $U^L_g \bar{U}^R_g$ for $g \in G$. This definition of SWSSB can therefore be interpreted in the doubled Hilbert space as the ordinary SSB of $G^L \times G^R$ down to a residual $G^{\text{diag}}$ symmetry \cite{lee2023quantumcriticality}.

The R\'enyi-2 correlator is conceptually appealing and admits simple calculation methods in a variety of settings \cite{lee2023quantumcriticality,bao2023mixedstatetopologicalordererrorfield,fan2024diagnostics,sala2024spontaneousstrongsymmetrybreaking}. Unfortunately, it suffers from two severe drawbacks. First, it has been recently observed that LRO in $R_2$ cannot always be used to faithfully distinguish between different mixed-state phases \cite{lessa2024strongtoweakspontaneoussymmetrybreaking}. That is, there exist simple examples of states within the same mixed-state phase---defined either by two-way connectivity via symmetric short-depth channels \cite{ellison2024classificationmixedstatetopologicalorders,sang2024mixed,Coser2019classificationof} or by more refined notions involving the nondivergence of the Markov length \cite{sang2024stabilitymixedstatequantumphases}---where one state exhibits long-range R\'enyi-2 correlations while the other does not. This behavior is closely related to the fact that $R_2$ is a second-moment observable in $\rho$; by expressing $\rho = \sum_i p_i \dyad{\psi_i}$ as an incoherent mixture of pure states, Eq.~\eqref{eq:renyi_2} effectively samples the states $\ket{\psi_i}$ with the modified probabilities $p_i^2 / \sum_i p_i^2$, biasing the observable toward the most probable states in the mixture.

Second, $R_2$ is inherently difficult to observe in an experiment, even when the density matrix $\rho$ is efficiently preparable. There exist several methods of measuring the purity $\tr \rho^2$ in the denominator or the inner product $\tr[\sigma \rho]$ of two density matrices in the numerator, such as the SWAP test \cite{ekertDirectEstimationsLinear2002} or classical shadow tomography \cite{huangPredictingManyProperties2020,elbenMixedStateEntanglementLocal2020,elbenRandomizedMeasurementToolbox2023}. However, since both of these quantities are exponentially small in the system size for generic mixed states, the sample complexity of estimating these quantities with statistical error comparable to their means is exponentially large. Even worse, the ratio of these two exponentially small quantities will exhibit a scaled statistical error proportional to their inverse means. One therefore expects that \textit{any} protocol for measuring $R_2$ will require exponentially many measurements.

A more robust theoretical measure of SWSSB is the fidelity correlator $F(x,y)$ proposed in Ref.~\cite{lessa2024strongtoweakspontaneoussymmetrybreaking}, given by the fidelity between the density matrices $\rho$ and $O_{xy} \rho O^{\dag}_{xy}$ \cite{footnote_3}:
\begin{equation}
\label{eq:fidelity}
	F(x,y) \coloneqq \tr \sqrt{ \sqrt{\rho} O_{xy} \rho O^{\dag}_{xy} \sqrt{\rho} } .
\end{equation}
The fidelity correlator has the physical interpretation of diagnosing the \textit{distinguishability} of $\rho$ and $O_{xy} \rho O^{\dag}_{xy}$, offering a natural mixed-state generalization of charge condensation interpretations of SSB in pure states. Most importantly, $F$ admits a \textit{stability theorem} \cite{lessa2024strongtoweakspontaneoussymmetrybreaking}: if $\rho$ exhibits SWSSB, in the sense that $F(x,y) \sim \mathcal{O}(1)$ as $\abs{x-y} \to \infty$, and $\mathcal{E}$ is a strongly symmetric finite-depth quantum channel, then $\mathcal{E}(\rho)$ also exhibits SWSSB. This means that any two states which are two-way connected by strongly symmetric finite-depth channels either both exhibit SWSSB or both do not exhibit SWSSB.

Unfortunately, $F(x,y)$ is also inherently difficult to measure in experiment, with no obvious measurement protocol beyond full-scale tomography. Additionally, the relation between LRO in $F(x,y)$ and the spontaneous breaking of a strong symmetry is \textit{a priori} unclear: as defined by $F$, there is no \textit{manifest} connection between SWSSB and ordinary SSB of a doubled symmetry group, as in the R\'enyi-2 definition.

\textit{R\'enyi-1 correlator.---} To alleviate these issues, we propose to define SWSSB using the ``R\'enyi-1'' correlator $R_1(x,y)$, which we define as
\begin{equation}
\label{eq:renyi-1}
		R_1(x,y) \coloneqq \tr[ O_{xy} \sqrt{\rho} O^{\dag}_{xy} \sqrt{\rho} ] = \bbra{\sqrt{\rho}} O_{xy}^L \bar{O}_{xy}^R \kett{\sqrt{\rho}} ,
\end{equation}
where $\kett{\sqrt{\rho}} \coloneqq (\sqrt{\rho} \otimes 1) \sum_{\vb{s}} \ket{\vb{s}}^L \ket{\vb{s}}^R$ is the \textit{canonical purification} (CP) of $\rho$ \cite{SOM}, which has the useful property that $\tr_R \kett{\sqrt{\rho}} \bbra{\sqrt{\rho}} = \rho$. As with the vectorized state $\kett{\rho}$, a density matrix $\rho$ with a strong $G$ symmetry gives rise to a CP state $\kett{\sqrt{\rho}}$ with $G^L \times G^R$ symmetry, such that $U_g^L \bar{U}_h^R \kett{\sqrt{\rho}} = e^{i(\varphi_g - \varphi_h)} \kett{\sqrt{\rho}}$. Similar in spirit to the R\'enyi-2 correlator \eqref{eq:renyi_2}, the R\'enyi-1 correlator $R_1$ defines SWSSB in a strongly symmetric mixed state $\rho$ as the conventional SSB of $G^L \times G^R$ symmetry in $\kett{\sqrt{\rho}}$ down to a residual $G^{\text{diag}}$ symmetry.

Despite the aesthetic similarities to $R_2$, $R_1$ also exhibits many close relations to the fidelity correlator \eqref{eq:fidelity} and the other information-theoretic observables for SWSSB proposed in Ref.~\cite{lessa2024strongtoweakspontaneoussymmetrybreaking}. For example, $R_1$ satisfies a quantum data processing inequality \cite{petzQuasientropiesStatesNeumann1985,wildeRecoverabilityHolevosJustGood2018} and a version of the Fuchs-van de Graaf inequality \cite{kholevoQuasiequivalenceLocallyNormal1972}. Most pertinently, we show in the Supplemental Material \cite{SOM} that $R_1$ is \textit{exactly} equal to $F$ for a large class of density matrices of interest: namely, stabilizer states affected by Pauli decoherence channels \cite{fan2024diagnostics,bao2023mixedstatetopologicalordererrorfield,lee2023quantumcriticality,sala2024spontaneousstrongsymmetrybreaking}, and thermal Gibbs states of stabilizer code Hamiltonians \cite{weinstein2019universality,lu2019singularity,lu2020detecting,li2024phase}. More generally, $R_1$ can be both upper and lower bounded by $F$ \cite{SOM}:
\begin{equation}
	[F(x,y)]^2 \leq R_1(x,y) \leq F(x,y) .
\end{equation}
Thus, SWSSB as defined by LRO in $R_1$ and $F$ are equivalent, and the former immediately inherits the stability theorem proven for the latter in Ref.~\cite{lessa2024strongtoweakspontaneoussymmetrybreaking}.

In addition to the theoretical and aesthetic benefits of the R\'enyi-1 correlator mentioned above, the final expression of Eq.~\eqref{eq:renyi-1} suggests a direct method for potential experimental detection of SWSSB: if the CP state $\kett{\sqrt{\rho}}$ can be prepared efficiently, then $R_1$ can easily be measured in this state as an ordinary two-point correlation function. LRO in $R_1$ in the absence of LRO in $\tr[ O_{xy} \rho] = \bbra{\sqrt{\rho}} O_{xy}^L \kett{\sqrt{\rho}}$ then immediately implies that the reduced density matrix on the ``left'' subsystem exhibits SWSSB.

\textit{Preparing canonical purifications.---} We now provide several examples of CP states exhibiting SWSSB which can be efficiently prepared in a quantum simulator. For clarity and technical simplicity, we focus on examples of SWSSB in mixed states with strong $\mathbb{Z}_2$ symmetry.

Each example consists of a system of $N$ qubits with Pauli operators $Z_j$ and $X_j$, arranged in a $d$-dimensional square lattice with periodic boundary conditions for concreteness. The initial state $\rho_0 = \dyad{+}^{\otimes N}$ is a pure product state which is $\mathbb{Z}_2$ symmetric under the parity operator $\Pi \coloneqq \prod_j X_j$. Trivially, its CP $\kett{\sqrt{\rho_0}}$ is a pure product state on the doubled system $LR$, which is separately symmetric under the ``left'' parity $\Pi^L = \prod_j X_j^L$ and the ``right'' parity $\Pi^R = \prod_j X_j^R$. In the examples to follow, the CP $\kett{\sqrt{\rho}}$ of another density matrix $\rho$ which exhibits SWSSB will spontaneously break these left and right parity symmetries, while remaining symmetric under the combined symmetry $\Pi^L \Pi^R$.

For our first example, we consider the effect of measuring $Z_i Z_j$ on each nearest-neighbor link $\expval{ij}$ of the lattice and discarding the measurement outcome; i.e., we subject each nearest-neighbor link to the dephasing channel $\mathcal{E}_{ij}(\rho) = \frac{1}{2}(\rho + Z_i Z_j \rho Z_i Z_j)$. The resulting density matrix $\rho_{\Pi} \coloneqq (1 + \Pi)/2^N$, which is an equally weighted incoherent mixture of all parity-even states, is a paradigmatic example of a mixed state exhibiting $\mathbb{Z}_2$ SWSSB \cite{ma2024topologicalphasesaveragesymmetries,lessa2024strongtoweakspontaneoussymmetrybreaking}. Specifically, while $\tr[Z_x Z_y \rho_{\Pi}] = 0$ for all $x \neq y$, all three of the observables \eqref{eq:renyi_2}--\eqref{eq:renyi-1} are unity for $O_{xy} = Z_x Z_y$.

One easily verifies that the canonical purification of $\rho_{\Pi}$ is
\begin{equation}
\label{eq:rho_Pi_CP}
	\kett{\sqrt{\rho_{\Pi}}} = 2^{(N-1)/2} \prod_{\expval{ij}} \qty( \frac{1 + Z_i^L Z_i^R Z_j^L Z_j^R}{2} ) \kett{\sqrt{\rho_0}} .
\end{equation}
That is, $\kett{\sqrt{\rho_{\Pi}}}$ is a stabilizer state with the $2N$ stabilizer generators $X_j^L X_j^R$, $Z_i^L Z_i^R Z_j^L Z_j^R$, and $\Pi^L$. The easiest method of preparing the state \eqref{eq:rho_Pi_CP} is to first prepare the state $\ket{\text{GHZ}}^L \ket{+}^R$, where $\ket{\text{GHZ}} \equiv \frac{1}{\sqrt{2}} (\ket{0}^{\otimes N} + \ket{1}^{\otimes N})$ is the $N$-qubit GHZ state and $\ket{+} \equiv \prod_j \ket{+}_j$, and then performing CNOT gates $CX^{R \to L}_j \equiv \frac{1}{2}(1 + X^L_j + Z^R_j - X^L_j Z^R_j)$ from each $j$th $R$ qubit to the corresponding $L$ qubit. The initial GHZ state can be prepared either in finite depth using $Z^L_i Z^L_j$ projective measurements and feedback \cite{tantivasadakarn2024long,lee2022decodingmeasurementpreparedquantumphases,zhu2023nishimori} or in depth $N$ via a sequential circuit of CNOT gates. Alternatively, one can directly prepare \eqref{eq:rho_Pi_CP} via projective measurements of $Z^L_i Z^R_i Z^L_j Z^R_j$ and feedback, using an identical strategy as for the GHZ state.

The facility of preparing the CP state $\kett{\sqrt{\rho_{\Pi}}}$ was due to its stabilizer nature. For our second example, we consider a generalization of the preceding construction: rather than dephasing every link $\expval{ij}$, we \textit{randomly} dephase each link with probability $p$ \cite{kunoStrongweakSymmetryBreaking2024}. This results in a statistical ensemble of stabilizer mixed states $\rho_{\ell} \coloneqq [ \prod_{\expval{ij} \in \ell} \mathcal{E}_{ij} ](\rho_0)$ with corresponding probabilities $p_{\ell} = (1-p)^{dN - \abs{\ell}} p^{\abs{\ell}}$, where $\ell$ denotes the subset of dephased links and $\abs{\ell}$ is the number of such links. 

To understand the structure of the states $\rho_{\ell}$, consider first the effect of $\mathcal{E}_{ij}$ on the initial state $\rho_0$. Initially, the stabilizer group of $\rho_0$ is generated by $\qty{X_j}_{j = 1}^N$. Upon applying $\mathcal{E}_{ij}$, the individual stabilizers $X_i$ and $X_j$ are eliminated from the stabilizer group, but the stabilizer $X_i X_j$ remains. More generally, if $\rho$ is a stabilizer state generated by a set of subregion parity operators $\Pi_{\mathcal{R}} \coloneqq \prod_{j \in \mathcal{R}} X_j$, the channel $\mathcal{E}_{ij}$ acting entirely within a region $\mathcal{R}$ leaves $\rho$ unaffected; conversely, if sites $i$ and $j$ belong to two distinct regions $\mathcal{R}$ and $\mathcal{R}'$, then the individual generators $\Pi_{\mathcal{R}}$ and $\Pi_{\mathcal{R}'}$ are eliminated and replaced with their product $\Pi_{\mathcal{R} \cup \mathcal{R}'}$. Thus, each random sampling of dephased links $\ell$ yields a percolation configuration, i.e., a collection $\mathcal{P}(\ell)$ of connected clusters of sites $\mathcal{R}$, and $\rho_{\ell}$ is the stabilizer state generated by the subregion parities $\Pi_{\mathcal{R}}$ for each cluster $\mathcal{R} \in \mathcal{P}(\ell)$. Explicitly, $\rho_{\ell} = \frac{1}{2^N} \prod_{\mathcal{R} \in \mathcal{P}(\ell)} \qty( 1 + \Pi_R )$. It is easy to check that $\tr[Z_x Z_y \rho_{\ell}] = 0$ for all $x \neq y$, but $R_1(x,y) = 1$ whenever $x$ and $y$ belong to the same region $\mathcal{R}$. Thus, the statistical ensemble of states $\rho_{\ell}$ undergoes an SWSSB transition for $d \geq 2$ in the universality class of $d$-dimensional bond percolation \cite{kunoStrongweakSymmetryBreaking2024,staufferIntroductionPercolationTheory2018}.

The CP states $\kett{\sqrt{\rho_{\ell}}}$ can be prepared in much the same way as Eq.~\eqref{eq:rho_Pi_CP}: given a disorder realization $\ell$, we simply apply the preceding protocol in each percolation cluster $R$. Bond configurations $\ell$ can be easily sampled, and the number of clusters is at most of order $N$, so it is not substantially more difficult to prepare any given state $\kett{\sqrt{\rho_{\ell}}}$ than to prepare $\kett{\sqrt{\rho_{\Pi}}}$.

The general strategy of promoting a pure SSB state on the $L$ subsystem to a CP state with SWSSB works for a large class of nonstabilizer states as well. For our third example, let $H \coloneqq - \sum_{\expval{ij}} Z_i Z_j - g \sum_j X_j$ be the Hamiltonian of the $d$-dimensional transverse-field Ising model (TFIM), and let $\ket{\psi_g}$ be its parity-even ground state as a function of the transverse field $g$. Notably, the wavefunction coefficients $\bra{\qty{x}} \ket{\psi_g}$ of $\ket{\psi_g}$ in the $X$ basis are nonnegative for all $g \geq 0$ \cite{auerbachInteractingElectronsQuantum1994}. As a result, the state 
\begin{equation}
	\begin{split}
		\kett{\Psi_g} &\coloneqq \prod_j CX_j^{R \to L} \ket{\psi_g}^L \ket{+}^R \\
		&= \sum_{\qty{x}} \bra{\qty{x}} \ket{\psi_g} \ket{\qty{x}}^L \ket{\qty{x}}^R
	\end{split}
\end{equation}
can be regarded as the CP of the mixed state $\rho_g \coloneqq \sum_{\qty{x}} \bra{\qty{x}} \ket{\psi_g}^2 \dyad{\qty{x}}$. Moreover, $\bbra{\Psi_g} Z^L_x Z^L_y \kett{\Psi_g}$ vanishes for $x \neq y$, while $\bbra{\Psi_g} Z^L_x Z^R_x Z^L_y Z^R_y \kett{\Psi_g} = \bra{\psi_g} Z_x Z_y \ket{\psi_g}$. In other words, the reduced density matrix $\rho_g$ exhibits SWSSB in the ferromagnetic phase of $H$. More generally, any ``sign-free'' wavefunction on $L$, including the ground state of any ``stoquastic'' Hamiltonian \cite{bravyi2007complexitystoquasticlocalhamiltonian}, can be trivially converted to a CP state on $LR$ by the same strategy. The reduced density matrices of the resulting states can then be used to construct a broad class of SWSSB states and transitions.

Since $\ket{\psi_g}$ is the ground state of a local gapped Hamiltonian, it can be efficiently prepared using adiabatic state preparation \cite{farhi2000quantumcomputationadiabaticevolution,das2008colloquium,georgscu2014quantum,albash2018adiabatic}. States in the paramagnetic phase are most easily achieved starting from the $g \to \infty$ ground state, while states in the ferromagnetic phase can be achieved by first constructing $\ket{\psi_{g = 0}} = \ket{\text{GHZ}}$ and then adiabatically turning on $g$.

As a final example, we consider the thermal Gibbs state of the TFIM, with Hamiltonian $H$ as above. Restricting to the parity-even sector via the projector $P_{\Pi} = \qty(\frac{1+\Pi}{2})$, the density matrix is $\rho_{\beta} \coloneqq P_{\Pi} e^{-\beta H} / \tr[ P_{\Pi} e^{-\beta H} ]$. The canonical purifications of thermal states are well known as thermofield double (TFD) states \cite{takahashiThermoFieldDynamics1996,maldacenaEternalBlackHoles2003,maldacenaCoolHorizonsEntangled2013,maldacena2018eternaltraversablewormhole}; in the present context, $\kett{\sqrt{\rho_{\beta}}}$ can be formally written using imaginary time evolution on $\kett{\sqrt{\rho_{\Pi}}}$:
\begin{equation}
\label{eq:tfd}
	\kett{\sqrt{\rho_{\beta}}} = \frac{1}{\sqrt{\tr[ P_{\Pi} e^{-\beta H}]}} \qty( e^{-\beta H / 2} \otimes 1 ) \kett{\sqrt{\rho_{\Pi}}} .
\end{equation}
At any finite temperature $\beta < \infty$ in $d = 1$, or above a critical temperature $\beta < \beta_c(g)$ in $d \geq 2$, the thermal density matrix $\rho_{\beta}$ exhibits short-range ferromagnetic correlations: $\tr[Z_x Z_y \rho_{\beta}] \to 0$ as $\abs{x-y} \to \infty$. It has been previously argued that such a finite-temperature paramagnetic phase generically exhibits SWSSB \cite{lessa2024strongtoweakspontaneoussymmetrybreaking}. This can be easily established in the extreme limits $g \to 0$ and $g \to \infty$, where the fidelity and R\'enyi-1 correlators are equal and can be computed exactly. In the Supplemental Material \cite{SOM}, we numerically demonstrate that the one-dimensional TFIM indeed exhibits SWSSB at all nonzero temperatures.

Here, rather than providing a specific algorithm for preparing $\kett{\sqrt{\rho_{\beta}}}$, we simply mention that several variational algorithms for preparing TFD states have already been proposed \cite{wu2019variational,cottrellHowBuildThermofield2019,martyn2019product,su2021variational,zhuGenerationThermofieldDouble2020}. A minor technical distinction between $\kett{\sqrt{\rho_{\beta}}}$ and the TFD states considered in these previous works is the restriction to the parity-even sector. As Eq.~\eqref{eq:tfd} shows, one simply needs to start their variational algorithm from the state $\kett{\sqrt{\rho_{\Pi}}}$ rather than a maximally entangled state on $LR$.

We end by mentioning a notable omission in this discussion: namely, the decohered quantum Ising model, a well-studied family of locally decohered states which exhibit an SWSSB transition \cite{lee2023quantumcriticality,lessa2024strongtoweakspontaneoussymmetrybreaking,sala2024spontaneousstrongsymmetrybreaking}. These mixed states $\rho_p$ are obtained from $\rho_0$ by applying the continuously tunable dephasing channel $\mathcal{E}_{ij}^p(\rho) = (1-p) \rho + p Z_i Z_j \rho Z_i Z_j$ to each nearest-neighbor link. In $d = 2$, $\rho_p$ exhibits an SWSSB transition in the universality class of the random-bond Ising model on the Nishimori line, while in $d = 3$ there is an SWSSB transition in the universality class of the random-plaquette $\mathbb{Z}_2$ gauge theory. As reviewed in the Supplemental Material \cite{SOM}, the R\'enyi-1 and fidelity correlators can be directly related to disorder parameter correlations in these effective statistical physics models, \textit{without} introducing additional replicas.

While it is trivial to construct a purification of $\rho_p$ using a Stinespring dilation \cite{Nielsen_Chuang_2010} of $\mathcal{E}_{ij}^p$, it is presently unclear how to transform this purification into the \textit{canonical} purification. Such a transformation can be achieved by a deep unitary circuit on the ancillary qubits alone. It is an interesting challenge for future work to find an explicit protocol for preparing the CPs of locally decohered mixed states such as $\rho_p$.

\textit{Discussion.---} We have proposed the R\'enyi-1 correlator $R_1(x,y)$ as a diagnostic of SWSSB with several theoretically and practically useful properties. Similar to its R\'enyi-2 counterpart $R_2$, $R_1$ defines SWSSB in a mixed state $\rho$ with $G$ symmetry as the spontaneous breaking of a doubled symmetry group $G^L \times G^R$ down to a diagonal subgroup $G^{\text{diag}}$. However, unlike $R_2$, LRO in $R_1$ is robust under strongly symmetric finite-depth quantum channels due to the stability theorem it inherits from the fidelity correlator \cite{lessa2024strongtoweakspontaneoussymmetrybreaking}. Moreover, the structure of $R_1$ as a two-point correlation function in the CP state $\kett{\sqrt{\rho}}$ suggests a simple protocol for observing a large class of SWSSB phases in experiment: if such a CP state can be efficiently prepared, and it is shown to exhibit the aforementioned pattern of symmetry breaking, then its reduced density matrix necessarily exhibits SWSSB. 

We have provided several examples of efficiently preparable CP states whose reduced density matrices exhibit SWSSB. For simplicity, we have restricted ourselves to CP states which are easily prepared using well-known state preparation protocols, and whose reduced density matrices are \textit{analytically} known to exhibit SWSSB. Experimentally, it would be far more interesting to investigate mixed states and CP states which are efficiently preparable, but nevertheless do not admit a simple theoretical understanding. Toward this end, a compelling future direction is the design of bespoke quantum algorithms for the preparation of CP states, potentially based on adiabatic evolution or using variational algorithms. 



Our method for observing SWSSB in a mixed state $\rho$ requires access to a particular purification of $\rho$. Ideally, one would prefer a protocol for detecting SWSSB which only requires access to $\rho$ itself. That is, given the ability to efficiently and repeatedly prepare an initially unknown mixed state $\rho$, one would like either a set of observables or perhaps a classical or quantum algorithm which can efficiently determine if $\rho$ exhibits SWSSB. Since there are likely to exist many efficiently preparable mixed states whose CPs are not efficiently preparable, such a protocol would be a key step towards determining the practical physical implications of SWSSB in real quantum platforms.


\begin{acknowledgments}
\textit{Acknowledgements.---} We thank Sajant Anand, Tyler Ellison, Tim Hsieh, Leonardo Lessa, Olumakinde Ogunnaike, Subhayan Sahu, Pablo Sala, and Chong Wang for insightful discussions, and especially Ehud Altman, Sam Garratt, and Zijian Wang for insightful discussions and helpful feedback on the manuscript. We also thank Jian-Hao Zhang for pointing out the connection to Holevo's just-as-good fidelity \cite{SOM}. This material is based upon work supported by the U.S. Department of Energy, Office of Science, National Quantum Information Science Research Centers, Quantum Systems Accelerator.

\textit{Note added.---} During the completion of this work, a preprint appeared \cite{liu2024diagnosingstrongtoweaksymmetrybreaking} which also discusses the R\'enyi-1 correlator in the context of SWSSB. Our results agree where they overlap.
\end{acknowledgments}

\bibliographystyle{apsrev4-2-author-truncate} 
\bibliography{refs}

\begin{thebibliography}{97}
\bibitem [{foo({\natexlab{b}})}]{footnote_2}%
  \BibitemOpen
  \href@noop {} {} \bibinfo {note} {In the case of a non-abelian symmetry, we have a multiplet of operators $O^a_x$ which transform in a higher dimensional representation of $G$: $U_g^{\dag} O^a_x U_g = \sum_b M^{ab}_g O^b_x$, where $M^{ab}_g$ is a unitary matrix representation of $g \in G$. One should then replace $O^L_x \bar{O}^R_x$ with $\sum_a [O^a_x]^L [\bar{O}^a_x]^R$ to obtain an order parameter which transforms nontrivially under $G^L$ and $G^R$ but remains invariant under $G^{\text{diag}}$.}\BibitemShut {Stop}%
\bibitem [{foo({\natexlab{c}})}]{footnote_3}%
  \BibitemOpen
  \href@noop {} {} \bibinfo {note} {We assume for simplicity throughout that $O_{xy}$ is unitary, such as when $O_x$ and $O_y$ are Pauli operators, so that $O_{xy} \rho O_{xy}^{\dag}$ is a valid density matrix. This assumption can easily be relaxed.}\BibitemShut {Stop}%
\end{thebibliography}


\onecolumngrid

\renewcommand{\thefigure}{S\arabic{figure}}
\renewcommand{\theequation}{S\arabic{equation}}
\renewcommand{\thetable}{S\Roman{table}}
\renewcommand{\thesection}{S\Roman{section}}

\setcounter{secnumdepth}{2}
\setcounter{equation}{0}

\newpage

\centerline{\large{\textbf{Supplemental Material For: Efficient Detection of Strong-To-Weak Spontaneous}}}
\vspace{0.1cm}
\centerline{\large{\textbf{Symmetry Breaking via the R\'enyi-1 Correlator}}}
\vspace{0.5cm}

\centerline{Zack Weinstein}

\centerline{\textit{Department of Physics, University of California, Berkeley, CA 94720, USA}}

\centerline{(Dated: \today)}

\tableofcontents

\section{Canonical Purification States}
\label{sec:CPstates}
In this Appendix, we provide a brief overview of canonical purification (CP) states. After explaining general features of these states, we specialize to describing the CPs of two particular cases of physical interest: namely, stabilizer mixed states, and Gibbs states.

Let $\rho = \sum_i p_i \dyad{\psi_i}$ be a density matrix of a quantum system, written without loss of generality as an incoherent mixture of orthonormal pure states $\ket{\psi_i}$. As its name suggests, the CP of $\rho$ is a particular purification of $\rho$, obtained as follows. First, since $\rho$ is nonnegative, its square root $\sqrt{\rho} = \sum_i \sqrt{p_i} \dyad{\psi_i}$ is a well-defined nonnegative Hermitian operator. Then, labeling the original system as $L$ and introducing an identical copy of the system labeled $R$ (`left' and `right' respectively), we obtain the CP state $\kett{\sqrt{\rho}}$ by acting $\sqrt{\rho}$ on the left half of an (unnormalized) maximally entangled state:
\begin{equation}
	\kett{\sqrt{\rho}} \define \qty( \sqrt{\rho} \otimes 1 ) \sum_{\vb{s}} \ket{\vb{s}}^L \ket{\vb{s}}^R = \sum_i \sqrt{p_i} \ket{\psi_i}^L \ket{\psi_i^*}^R ,
\end{equation}
where $\ket{\vb{s}}$ is a complete orthonormal basis of the original system, and the state $\ket{\psi_i^*}$ has complex-conjugated matrix elements $\bra{\vb{s}} \ket{\psi_i^*} = \bra{\vb{s}} \ket{\psi_i}^*$ relative to $\ket{\psi_i}$. We use kets with doubled angular brackets $\kett{\cdot}$ to denote states in the doubled Hilbert space $LR$, kets with single brackets and a superscript $\ket{\cdot}^{L,R}$ to denote states in the left or right Hilbert space, and single brackets without a superscript $\ket{\cdot}$ to denote states in the original Hilbert space.

Despite its name, the canonical purification $\kett{\sqrt{\rho}}$ is not entirely unique; it requires a particular choice of basis $\ket{\vb{s}}$ of the original Hilbert space, and different choices of basis need not yield the same state $\kett{\sqrt{\rho}}$. In particular, the complex conjugation operation required to obtain $\ket{\psi_i^*}$ from $\ket{\psi_i}$ is basis-dependent: complex conjugation of wavefunction coefficients in two different bases need not agree. The canonical purifications constructed with two different sets of bases $\ket{\vb{r}}$ and $\ket{\vb{s}}$ agree if they are related by an \textit{orthogonal} transformation, i.e., $\ket{\vb{r}} = \sum_{\vb{s}} \mathsf{O}_{\vb{r} \vb{s}} \ket{\vb{s}}$ with $\sum_{\vb{r}} \mathsf{O}_{\vb{r} \vb{s}} \mathsf{O}_{\vb{r} \vb{s}'} = \delta_{\vb{s} \vb{s}'}$. For example, in a system of qubits, canonical purifications defined with respect to the Pauli-$Z$ eigenbasis and the Pauli-$X$ eigenbasis are equivalent. Throughout this work, we assume that the basis $\ket{\vb{s}}$ is the computational basis, i.e., the Pauli-$Z$ eigenbasis.

Any two purifications of a quantum mixed state are equivalent up to an isometry $V$ on the auxiliary system; this follows easily from the uniqueness properties of the Schmidt decomposition \cite{Nielsen_Chuang_2010}. In other words, an arbitrary purification $\kett{\Psi_{\rho}}$ of $\rho$ can be written as
\begin{equation}
	\kett{\Psi_{\rho}}^{LA} = (1 \otimes V^{A \leftarrow R}) \kett{\sqrt{\rho}} = \qty( \sqrt{\rho} \otimes V^{A \leftarrow R} ) \sum_{\vb{s}} \ket{\vb{s}}^L \ket{\vb{s}}^R .
\end{equation}
Thus, \textit{any} purification of $\rho$ can be brought to the form of a canonical purification by an isometry acting on the auxiliary space $A$ alone. Note that the dimension of the auxiliary Hilbert space $A$ must be at least as large as the rank of $\rho$. Therefore, for a generic density matrix of full rank, the auxiliary system $A$ must be at least as large as $R$. 

Naturally, since $\kett{\sqrt{\rho}}$ is a purification of $\rho$, expectation values of observables $O^L \define (O \otimes 1)$ in the left subsystem alone reproduce expectation values of $\rho$: $\bbra{\sqrt{\rho}} O^L \kett{\sqrt{\rho}} = \tr[O \rho]$. Meanwhile, expectation values of observables $\bar{O}^R \define (1 \otimes O^*)$ yield $\bbra{\sqrt{\rho}} \bar{O}^R \kett{\sqrt{\rho}} = \tr[\rho O^{\dag}]$. Finally, general two-sided expectation values can be written in terms of a trace as $\bbra{\sqrt{\rho}} O_1^L \bar{O}_2^R \kett{\sqrt{\rho}} = \tr[O_1 \sqrt{\rho} O_2^{\dag} \sqrt{\rho} ]$. Each of these identities can be derived by simple index-chasing, or more rapidly via tensor network diagrams.

We now discuss two broad classes of canonical purification states with insightful formal expressions. 

\subsection{Thermofield Double States}
First, we consider the canonical purification of a thermal state $\rho_{\beta} \define e^{-\beta H} / \mathcal{Z}_{\beta}$, where $\mathcal{Z}_{\beta} \define \tr e^{-\beta H}$ is the partition function of the Hamiltonian $H$ at inverse temperature $\beta$. In this setting, the canonical purification of $\rho_{\beta}$ is referred to as the thermofield double (TFD) state. It is frequently written in the following various forms:
\begin{equation}
	\begin{split}
		\kett{\sqrt{\rho_{\beta}}} &= \frac{1}{\sqrt{\mathcal{Z}_{\beta}}} \qty( e^{-\beta H/2} \otimes 1 ) \sum_{\vb{s}} \ket{\vb{s}}^L \ket{\vb{s}}^R \\
		&= \frac{1}{\sqrt{\mathcal{Z}_{\beta}}} \qty( e^{-\beta H/4} \otimes e^{-\beta H^T/4} ) \sum_{\vb{s}} \ket{\vb{s}}^L \ket{\vb{s}}^R \\
		&= \frac{1}{\sqrt{\mathcal{Z}_{\beta}}} \sum_n e^{-\beta E_n / 2} \ket{E_n}^L \ket{E_n^*}^R ,
	\end{split}\end{equation}
where $\ket{E_n}$ are the energy eigenstates of $H$ with the energies $E_n$. In the middle expression, we have split up the factor of $e^{-\beta H / 2}$ across the two copies $L$ and $R$ of the system, which requires replacing $H$ with $H^T$ in the right subsystem. Since $H$ is Hermitian, we could equivalently write $H^* = H^T$, where $H^*$ is the matrix with complex conjugated matrix elements in the basis $\ket{\vb{s}}$. Note that two-sided correlation functions, such as the R\'enyi-1 correlator [Eq.~\eqref{eq:renyi_1}], have a particularly simple interpretation in the TFD state $\kett{\sqrt{\rho_{\beta}}}$: they are simply imaginary time-ordered correlation functions, with one operator evolved to imaginary time $\tau = \beta / 2$:
\begin{equation}
	\begin{split}
		\bbra{\sqrt{\rho_{\beta}}} O^L_1 \bar{O}^R_2 \kett{\sqrt{\rho_{\beta}}} &= \frac{1}{\mathcal{Z}_{\beta}} \tr \qty[ e^{-\beta H/2} O_1 e^{-\beta H/2} O_2^{\dag} ] \\
		&= \frac{1}{\mathcal{Z}_{\beta}} \tr \qty[ e^{-\beta H} \qty( e^{\beta H/2} O_1 e^{-\beta H/2} ) O_2^{\dag} ] \\
		&= \expval{ O_1(\tau = \beta / 2) O_2^{\dag}(\tau = 0) }_{\beta} .
	\end{split}
\end{equation}

\subsection{Canonical Purifications of Stabilizer States}
Second, we consider the case of a stabilizer state $\rho_{\mathcal{S}}$. Let $\mathcal{S}$ be an $[[N,k]]$ stabilizer code on $N$ physical qubits with $N-k$ generators $\qty{g_a}_{a = 1}^{N-k}$, and $2k$ logical operators $\qty{\mathcal{Z}_n , \mathcal{X}_n}_{n = 1}^k$. All of the generators commute amongst each other and with all of the logical operators, while the logical operators form a Pauli algebra of $k$ qubits: 
\begin{equation}
	\comm{g_a}{g_b} = 0, \quad \comm{g_a}{\mathcal{Z}_n} = \comm{g_a}{\mathcal{X}_n} = 0, \quad \comm{\mathcal{Z}_n}{\mathcal{Z}_m} = \comm{\mathcal{X}_n}{\mathcal{X}_m} = \comm{\mathcal{Z}_n}{\mathcal{X}_m} = 0 \text{ for } n \neq m, \quad \acomm{\mathcal{Z}_n}{\mathcal{X}_n} =0 .
\end{equation}
The stabilizer state $\rho_{\mathcal{S}}$ is defined as the uniform projector onto the $2^k$-dimensional stabilizer subspace $V_{\mathcal{S}} = \qty{\ket{\psi} : g \ket{\psi} = \ket{\psi}, g \in \mathcal{S}}$. It can be written explicitly in terms of the generators $g_a$ as
\begin{equation}
\label{eq:stab_state}
	\rho_{\mathcal{S}} = \frac{1}{2^k} \prod_{a = 1}^{N-k} \qty( \frac{1 + g_a}{2} ) .
\end{equation}
In this form, it is trivial to take the square root: $\sqrt{\rho_{\mathcal{S}}} = 2^{k/2} \rho_{\mathcal{S}}$. To obtain $\kett{\sqrt{\rho_{\mathcal{S}}}}$, we simply act $\sqrt{\rho_{\mathcal{S}}}$ on the left side of the maximally entangled state, $\kett{\Phi} \propto \sum_{\vb{s}} \ket{\vb{s}}^L \ket{\vb{s}}^R$. This state is itself a stabilizer state, with a stabilizer group generated by $X^L_j X^R_j$ and $Z^L_j Z^R_j$ for $j = 1, \ldots, N$. The projectors $\frac{1}{2}(1 + g_a^L)$ can then be understood as projective measurements of the stabilizers $g^L_a$ performed on the left system, with the postselected outcomes $+1$; this reorganizes the stabilizer group by eliminating all stabilizers which anticommute with $g_a^L$ and appending these generators to the generating set. To determine the resulting stabilizer group, it is convenient to first perform a change of basis on the generators of $\kett{\Phi}$. By inspection, we can choose to represent the stabilizer group $\mathcal{S}(\kett{\Phi})$ of $\kett{\Phi}$ with the following set of generators:
\begin{equation}
	\mathcal{S}(\kett{\Phi}) = \expval{ g_a^L \bar{g}_a^R, \quad h_a^L \bar{h}_a^R, \quad \mathcal{Z}_n^L \bar{\mathcal{Z}}_n^R, \quad \mathcal{X}_n^L \bar{\mathcal{X}}_n^R } ,
\end{equation}
where the $h_a$ operators are the `destabilizers' \cite{aaronsonImprovedSimulationStabilizer2004}, i.e., Pauli operators which satisfy $\acomm{g_a}{h_a} = 0$ and $\comm{g_a}{h_b} = 0$ for $a \neq b$, thereby completing the algebra $\qty{g_a, h_a, \mathcal{Z}_n, \mathcal{X}_n}$ into a full $N$-qubit Pauli algebra. After measuring each $g_a^L$, the generators $h_a^L \bar{h}_a^R$ are eliminated, and we have the following stabilizer group\footnote{To make contact with the expression for $\kett{\sqrt{\rho_{\Pi}}}$ provided in the main text, we use the single generator $\Pi$ for our stabilizer group, the single destabilizer $Z_1$, and the $2(N-1)$ logical operators $X_2, \ldots, X_N$ and $Z_1 Z_2, \ldots , Z_1 Z_N$. After the postselected measurement, the stabilizer group of the doubled state is generated by $\Pi^L$, $\Pi^R$, $X_j^L X_j^R$, and $Z_1^L Z_1^R Z_j^L Z_j^R$ for $j = 2, \ldots , N$.} for $\kett{\sqrt{\rho_{\mathcal{S}}}}$:
\begin{equation}
	\mathcal{S}(\kett{\sqrt{\rho_{\mathcal{S}}}}) = \expval{ g_a^L, \quad \bar{g}_a^R, \quad \mathcal{Z}_n^L \bar{\mathcal{Z}}_n^R, \quad \mathcal{X}_n^L \bar{\mathcal{X}}_n^R } .
\end{equation}
This prescription gives an easy recipe for constructing the canonical purification of any stabilizer state $\rho_{\mathcal{S}}$: introduce two copies of the system, one in the state $\rho_{\mathcal{S}}$ and the other in the state $\rho_{\mathcal{S}}^*$, and then maximally entangle their logical spaces by introducing the $2k$ additional stabilizers $\mathcal{Z}_n^L \bar{\mathcal{Z}}_n^R, \mathcal{X}_n^L \bar{\mathcal{X}}_n^R$.

\section{Comparison of R\'enyi-1 Correlator and Other Measures of SWSSB}
In this Appendix, we discuss several theoretically useful features of the R\'enyi-1 correlator $R_1(x,y)$, defined for local observables $O_x, O_y$ and a density matrix $\rho$ via
\begin{equation}
\label{eq:renyi_1}
	R_1(x,y) \define \tr \qty[ O_x O^{\dag}_y \sqrt{\rho} O_y O^{\dag}_x \sqrt{\rho} ] = \bbra{\sqrt{\rho}} O_x^L \bar{O}^R_x [O^L_y \bar{O}^R_y]^{\dag} \kett{\sqrt{\rho}} .
\end{equation}
In particular, we shall compare $R_1$ with several other recently proposed diagnostics of strong-to-weak spontaneous symmetry breaking (SWSSB): namely,
%
%
%
the fidelity correlator,
\begin{equation}
\label{eq:supp_fidelity}
	F(x,y) \define \tr \sqrt{ \sqrt{\rho} O_x O^{\dag}_y \rho O_y O^{\dag}_x \sqrt{\rho} } ,
\end{equation}
the relative entropy correlator,
\begin{equation}
\label{eq:rel_entropy}
	\mathcal{D}(x,y) \define \tr \qty{ \rho \qty[ \log \rho - \log( O_x O^{\dag}_y \rho O_y O^{\dag}_x ) ] },
\end{equation}
the trace distance correlator,
\begin{equation}
\label{eq:tr_dist}
	D_1(x,y) \define \frac{1}{2} \norm{ \rho - O_x O^{\dag}_y \rho O_y O^{\dag}_x }_1 ,
\end{equation}
and the R\'enyi-2 correlator,
\begin{equation}
\label{eq:supp_renyi_2}
	R_2(x,y) \define \frac{\tr \qty[ O_x O^{\dag}_y \rho O_y O^{\dag}_x \rho ] }{\tr \rho^2} = \frac{\bbra{\rho}  O_x^L \bar{O}^R_x [O^L_y \bar{O}^R_y]^{\dag} \kett{\rho} }{\bbrakett{\rho}{\rho}} .
\end{equation}
Eqs.~\eqref{eq:supp_fidelity}, \eqref{eq:rel_entropy}, and \eqref{eq:tr_dist}, proposed by Refs.~\cite{bao2023mixedstatetopologicalordererrorfield,fan2024diagnostics,lee2023quantumcriticality,lessa2024strongtoweakspontaneoussymmetrybreaking}, exhibit operationally meaningful information-theoretic interpretations, and the fidelity correlator in particular admits a useful ``stability theorem'' \cite{lessa2024strongtoweakspontaneoussymmetrybreaking}. The R\'enyi-2 correlator, which has received the most direct attention thus far \cite{zhang2024strongtoweakspontaneousbreaking1form,lee2023quantumcriticality,bao2023mixedstatetopologicalordererrorfield,fan2024diagnostics,sala2024spontaneousstrongsymmetrybreaking}, offers a particularly transparent interpretation of SWSSB in terms of spontaneously breaking ``left'' and ``right'' symmetries of a doubled system down to a diagonal subgroup; however, as we shall see below, it generically disagrees with the previously mentioned observables on whether a given mixed state exhibits SWSSB, and is expected to exhibit critical phenomena in an altogether different universality class. In particular, $R_2$ does not respect two-way connectivity, in the sense that a state with long-range R\'enyi-2 correlations and a state without such correlations can be two-way connected by finite-depth symmetric channels.

In contrast to all of these observables, the R\'enyi-1 correlator combines the best features of both the R\'enyi-2 correlator and the fidelity correlator: it admits similar symmetry-breaking interpretations to the R\'enyi-2 correlator, and it shares several useful information-theoretic properties with the fidelity correlator. Most pertinently for SWSSB, the R\'enyi-1 correlator inherits the stability theorem from the fidelity correlator due to the two-sided bound proven in Appendix~\ref{sec:bounds}.


First, we shall discuss several information-theoretic properties of the R\'enyi-1 correlator, which are highly analogous to those of the fidelity, relative entropy, and trace distance correlators discussed in Ref.~\cite{lessa2024strongtoweakspontaneoussymmetrybreaking}. Then, we will show that $R_1$ is exactly equal to the fidelity correlator for two large classes of mixed states of recent theoretical interest: namely, stabilizer states affected by Pauli decoherence channels, and thermal Gibbs states of stabilizer code Hamiltonians.

\subsection{Information-Theoretic Properties of the R\'enyi-1 Correlator}
Let us assume for simplicity that $O_x O_y^{\dag}$ is unitary, so that $\sigma \equiv O_x O_y^{\dag} \rho O_y O_x^{\dag}$ is a valid density matrix. Then, $R_1 = \tr[\sqrt{\sigma} \sqrt{\rho}]$ is a lesser-known distance measure between the two density matrices $\sigma$ and $\rho$ called Holveo's ``just-as-good'' fidelity\footnote{We thank Jian-Hao Zhang for pointing this out to us.} \cite{kholevoQuasiequivalenceLocallyNormal1972,wildeRecoverabilityHolevosJustGood2018}, or the ``quantum affinity'' \cite{luoInformationalDistanceQuantumstate2004}. Similar to the well-known distance measures in Eqs.~\eqref{eq:supp_fidelity}, \eqref{eq:rel_entropy}, and \eqref{eq:tr_dist}, the Holevo fidelity satisfies a quantum data processing inequality: if $\mathcal{E}$ is \textit{any} quantum channel, and $\sigma$ and $\rho$ are two arbitrary density matrices, then the following inequality holds \cite{wildeRecoverabilityHolevosJustGood2018}:
\begin{equation}
	\tr[ \sqrt{ \mathcal{E}(\sigma) } \sqrt{\mathcal{E}(\rho)} ] \geq \tr[ \sqrt{\sigma} \sqrt{\rho} ] .
\end{equation}
In other words, two density matrices can only become less distinguishable under the action of a quantum channel. In the special case where $O_x O^{\dag}_y$ commutes with all of the Kraus operators of a channel $\mathcal{E}$, then this observation is sufficient to prove that long-range order (LRO) in the R\'enyi-1 correlator [as well as in the correlators \eqref{eq:supp_fidelity}, \eqref{eq:rel_entropy}, \eqref{eq:tr_dist}] is stable under the channel $\mathcal{E}$. 

Additionally, $R_1$ satisfies an inequality which is completely analogous to the well-known Fuchs-van de Graaf inequality \cite{wildeRecoverabilityHolevosJustGood2018,Nielsen_Chuang_2010}: so long as $O_x O^{\dag}_y$ is unitary, we have the two-sided inequality
\begin{equation}
	1 - R_1(x,y) \leq D_1(x,y) \leq \sqrt{1 - [R_1(x,y)]^2} .
\end{equation}
This inequality is identical in form to the Fuchs-van de Graaf inequality, with the Uhlmann fidelity \eqref{eq:supp_fidelity} replaced by the Holevo fidelity $R_1$. As a consequence of both of these inequalities, SWSSB as defined by $R_1$, $D_1$, and $F$ are all equivalent. The inequalities also serve to provide a useful physical interpretation of $R_1$: its square serves as an upper bound for the probability of error in discriminating $\sigma$ from $\rho$ in a hypothesis testing experiment \cite{wildeRecoverabilityHolevosJustGood2018}. Thus, LRO in $R_1$ suggests that, when two distant symmetry charges are added to $\rho$, the resulting state $\sigma$ is \textit{indistinguishable} from $\rho$.

\subsection{Stabilizer States under Pauli Noise}
\label{subsec:stab_pauli}
Let $\mathcal{S} = \expval{ g_1, \ldots , g_{N-k} }$ be an $[[N,k]]$ stabilizer code on $N$ physical qubits as defined in Appendix \ref{sec:CPstates}. We initialize our $N$-qubit system in the stabilizer state $\rho_{\mathcal{S}}$ of $\mathcal{S}$, as in Eq.~\eqref{eq:stab_state}. The system is then subjected to a general Pauli channel $\mathcal{E}$, which applies the Pauli error $e$ with probability $p_e$:
\begin{equation}
\label{eq:pauli_channel}
	\rho = \mathcal{E}(\rho_{\mathcal{S}}) = \sum_{e \in \mathcal{P}_N} p_e e \rho_{\mathcal{S}} e^{\dag}, \quad \sum_{e \in \mathcal{P}_N} p_e = 1 , 
\end{equation}
where $\mathcal{P}_N$ is the group of $N$-qubit Pauli strings, together with phases $\pm 1, \pm i$. Note that each Pauli error $e \in \mathcal{P}_N$ is unitary, so that $\mathcal{E}$ is a proper quantum channel.

The number of distinct Pauli errors $e \in \mathcal{P}_N$ is exponentially large in $N$. Many of these errors act \textit{degenerately} on the state $\rho_{\mathcal{S}}$; two errors $e$ and $e'$ are called degenerate, and are grouped into the same equivalence class $s = [e]$, if $e \rho_{\mathcal{S}} e^{\dag} = e' \rho_{\mathcal{S}} e'^{\dag}$ . If $\mathcal{S}$ is regarded as a quantum error correcting code encoding $k$ logical qubits, then each equivalence class $s$ is a set of errors which yield the same syndrome measurement. We can partially simplify the state $\rho$ by writing it as a sum over a smaller (but still exponentially large) set of equivalence classes $s$:
\begin{equation}
\label{eq:pauli_channel_reraveling}
	\rho = \sum_s P_s \rho_s, \quad P_s = \sum_{e \in s} p_e, \quad \rho_s = e \rho_{\mathcal{S}} e^{\dag} \text{ for } e \in s .
\end{equation}
Crucially, the density matrices $\rho_s$ corresponding to particular syndrome measurements are orthogonal projectors, such that $\rho_s \rho_{s'} = \frac{1}{2^k} \delta_{s s'} \rho_s$. This follows from the original definition~\eqref{eq:stab_state} of $\rho_{\mathcal{S}}$ as a projector onto a stabilizer space, and the observation that a given Pauli error $e$ can only commute or anticommute with the generators $g_a$. As a result, each state $e \rho_{\mathcal{S}} e^{\dag}$ is itself a projector onto a stabilizer space, where we obtain the new stabilizer group by simply flipping the sign of each generator which anticommutes with $e$.

The above representation of $\rho$ remains purely formal. Nevertheless, many information-theoretic observables can be usefully expressed in this formal representation. For example, if $O$ is a general Pauli operator and $Os$ denotes the left coset\footnote{In other words, if $e$ is a representative of the equivalence class $s$ (i.e., $s = [e]$), then $Os \define [Oe]$.} of the equivalence class $s$ by $O$, then
\begin{equation}
\label{eq:P_Os}
	O \rho O^{\dag} = \sum_s P_s O \rho_s O^{\dag} = \sum_s P_s \rho_{Os} = \sum_s P_{Os} \rho_s .
\end{equation}
Additionally, the square root $\sqrt{\rho}$ and logarithm $\log \rho$ of $\rho$ can easily be computed, giving the results\footnote{Strictly speaking, the logarithm is not well-defined if $P_s = 0$ for some $s$ (equivalently, if the set of projectors $\rho_s$ with nonzero weight in $\rho$ do not form a complete set). In computing $\mathcal{D}(x,y)$, we set $P_s \log P_{s'} = 0$ if $P_s = P_s' = 0$, and $P_s \log P_{s'} = -\infty$ if $P_{s'} = 0 \neq P_s$.}
\begin{equation}
	\sqrt{\rho} = 2^{k/2} \sum_{s} \sqrt{P_s} \rho_s, \quad \log \rho = -k + 2^k \sum_s [\log P_s] \rho_s .
\end{equation}
Using these, we can immediately obtain formal expressions for a variety of information-theoretic probes of SWSSB. Letting $O \equiv O_x O^{\dag}_y$ in Eq.~\eqref{eq:renyi_1}, the R\'enyi-1 correlator is immediately given by
\begin{equation}
\label{eq:renyi_1_2}
	R_1(x,y) = \sum_s \sqrt{P_{Os} P_s} = \sum_s P_s \sqrt{ \frac{P_{Os}}{P_s} } = \sum_{e \in \mathcal{P}_N} p_e \sqrt{ \frac{P_{[Oe]}}{P_{[e]}} } ,
\end{equation}
where we have used the fact that $P_{[e]}$ depends only on the equivalence class $e$ to write $R_1$ as a sum over all Pauli errors. Similarly, the correlators of Eqs.~\eqref{eq:supp_fidelity},~\eqref{eq:rel_entropy},~\eqref{eq:tr_dist},~\eqref{eq:supp_renyi_2} can be formally written as
\begin{subequations}
	\begin{align}
		F(x,y) &= \sum_s \sqrt{P_{Os} P_s} = \sum_{e \in \mathcal{P}_N} p_e \sqrt{\frac{P_{[Oe]}}{P_{[e]}}}, \label{eq:fidelity_2} \\
		\mathcal{D}(x,y) &= - \sum_s P_s \log \qty{ \frac{P_{Os}}{P_s} } = - \sum_{e \in \mathcal{P}_N} p_e \log \qty{ \frac{P_{[Oe]}}{P_{[e]}} },  \label{eq:rel_entropy_2} \\
		D_1(x,y) &= \frac{1}{2} \sum_s \abs{P_{s} - P_{O s}} = \frac{1}{2} \sum_{e \in \mathcal{P}_N} p_e \abs{1 - \frac{P_{[Oe]}}{P_{[e]}}} , \label{eq:tr_dist_2} \\
		R_2(x,y) &= \frac{\sum_s P_s P_{Os}}{\sum_s P_s^2} = \frac{\sum_{e \in \mathcal{P}_N} p_e P_{[O e]}}{\sum_{e \in \mathcal{P}_N} p_e P_{[e]}} \label{eq:renyi_2_2} .
	\end{align}
\end{subequations}
Note in particular that $R_1(x,y)$ exactly agrees with the fidelity correlator $F(x,y)$.

In the context of locally decohered stabilizer states \cite{dennisTopologicalQuantumMemory2002,fan2024diagnostics,bao2023mixedstatetopologicalordererrorfield,lee2023quantumcriticality,lessa2024strongtoweakspontaneoussymmetrybreaking,sala2024spontaneousstrongsymmetrybreaking,ma2024topologicalphasesaveragesymmetries,chen2024separability,su2024tapestry,li2024replicatopologicalorderquantum,sala2024stabilityloopmodelsdecohering,li2024perturbativestabilityerrorcorrection,sriram2024nonuniformnoiseratesgriffiths,hauser2024informationdynamicsdecoheredquantum,Guo2024twodimensional,lee2024symmetryprotectedtopologicalphases,zhang2024quantumcommunicationmixedstateorder}, where the stabilizer generators $g_a$ are local and the Pauli channel $\mathcal{E}$ is a product of local error channels, the quantities $P_s$ can be understood as partition functions of disordered statistical mechanics models. Each possible error $e$ yields a disorder realization with probability $p_e$, but the partition function $P_{[e]}$ depends only on the equivalence class of $e$. The fact that $P_s = \sum_{e \in s} p_e$ is given by a sum over the disorder realization probabilities leads to a ``Nishimori''-like condition, i.e., a relation between parameters in the effective disordered statistical mechanics model and the parameters generating its disorder. In this interpretation, it is crucial to note that each of the quantities~\eqref{eq:renyi_1_2},~\eqref{eq:fidelity_2},~\eqref{eq:rel_entropy_2}, and~\eqref{eq:tr_dist_2} are ``quenched'' averages, while the R\'enyi-2 correlator~\eqref{eq:renyi_2_2} is given by an ``annealed'' average. This observation suggests that the first three observables are likely to all exhibit the same critical phenomena and provide equivalent probes of SWSSB in reasonably local stabilizer codes under local decoherence, while the R\'enyi-2 correlator generically exhibits a phase transition at a different location and with a different universality class. This picture is easily verified in the most well-known problems of stabilizer codes under local decoherence \cite{dennisTopologicalQuantumMemory2002,fan2024diagnostics,bao2023mixedstatetopologicalordererrorfield,lee2023quantumcriticality}.

\subsection{Gibbs States of Stabilizer Code Hamiltonians}
\label{subsec:gibbs}
Consider a system of $N$ qubits with a stabilizer code Hamiltonian of the form $H = - \sum_{a} g_{a}$, such that each $g_{a}$ is a Pauli string and $[g_a, g_b] = 0$. Typically the stabilizers $g_a$ in the Hamiltonian are \textit{not} algebraically independent, but are instead chosen as a set of local operators; we need not assume anything about the algebra or locality of operators in $H$ in this section. At finite inverse temperature $\beta < \infty$, the system is described by the following Gibbs state:
\begin{equation}
	\rho_{\beta} \define \frac{e^{-\beta H}}{\mathcal{Z}_{\beta}}, \quad \mathcal{Z}_{\beta} \define \tr e^{-\beta H} .
\end{equation}
In this state, the canonical purification $\kett{\sqrt{\rho_{\beta}}}$ is well-known as the \textit{thermofield double} (TFD) state. Such states can be prepared with relative efficiency in quantum simulators using variational quantum simulation \cite{wu2019variational,cottrellHowBuildThermofield2019,martyn2019product,su2021variational,zhuGenerationThermofieldDouble2020}, and in numerical simulations of tensor network states using time-evolving block decimation (TEBD) \cite{paeckelTimeevolutionMethodsMatrixproduct2019}. 

Once again, let us assume that $O \equiv O_x O^{\dag}_y$ is a Pauli operator for simplicity. We can then show  once again that the R\'enyi-1 correlator and the fidelity correlator are equal in the Gibbs state $\rho_{\beta}$. First note that if $O$ commutes with $H$, then $R_1(x,y) = F(x,y) = 1$. More generally, if $O$ anticommutes with a stabilizer $g_a$, then we have the following identity:
\begin{equation}
		O e^{\beta g_a / 2} O^{\dag} e^{\beta g_a / 2} = e^{\beta (O g_a O^{\dag}) / 2} e^{\beta g_a / 2} = e^{-\beta g_a / 2} e^{\beta g_a / 2} = 1.
\end{equation}
Using this observation, one easily computes both the R\'enyi-1 correlator and the fidelity correlator:
\begin{equation}
	R_1(x,y) = F(x,y) = \frac{1}{\mathcal{Z}_{\beta}} \tr \exp \qty{ \beta \sum_a J^O_a g_a }, \quad J^O_a = \begin{cases}
		1, & [O, g_a] = 0 \\
		0, & \acomm{O}{g_a} = 0
	\end{cases} .
\end{equation}
In other words, both observables can be understood as a type of disorder operator which eliminates stabilizers in $H$ which fail to commute with $O$. Formal expressions for the other three observables $\mathcal{D}(x,y)$, $D_1(x,y)$, and $R_2(x,y)$ can be computed similarly.

\section{Upper and Lower Bounds on the R\'enyi-1 Correlator}
\label{sec:bounds}
In this Appendix, we show that the R\'enyi-1 correlator $R_1(x,y)$ of an operator $O_x$ can be upper and lower bounded by the fidelity correlator $F(x,y)$ of the same operator. Ref.~\cite{lessa2024strongtoweakspontaneoussymmetrybreaking} proposed to define strong-to-weak spontaneous symmetry breaking (SWSSB) via long-range order (LRO) in the fidelity correlator $F(x,y)$ in the absence of LRO in ordinary correlators $\tr[O_x O^{\dag}_y \rho]$. The inequalities derived below [Eq.~\eqref{eq:fidelity_R1_inequality}] imply that long-range order in $F(x,y)$ and the R\'enyi-1 correlator $R_1(x,y)$ are equivalent, and can be used interchangeably to define SWSSB. Consequently, the R\'enyi-1 correlator immediately inherits the stability theorem proven for the fidelity correlator in Ref.~\cite{lessa2024strongtoweakspontaneoussymmetrybreaking}: if $\rho$ exhibits LRO in $R_1(x,y)$ [i.e., $R_1(x,y) \sim \mathcal{O}(1)$ as $\abs{x-y} \to \infty$], and $\mathcal{E}$ is a strongly symmetric finite-depth quantum channel, then $\mathcal{E}(\rho)$ also exhibits LRO\footnote{A minor subtlety is that, in general, the charged operator $O_x$ used in $R_1$ before the channel is applied may be different than the charged operator used in $R_1$ after the channel is applied.} in $R_1(x,y)$.

For convenience of notation, let $\sigma \equiv O_x O^{\dag}_y \rho O_y O_x^{\dag}$. We assume for simplicity that $O_x O^{\dag}_y$ is unitary (such as when $O_x$ and $O_y$ are Pauli operators) so that $\sigma$ is a valid density matrix, although this assumption can easily be relaxed. 

The upper bound on $R_1$ follows trivially from Uhlmann's theorem \cite{Nielsen_Chuang_2010}, which states that the fidelity between two density matrices $\sigma$ and $\rho$ is given by the supremum of all pure-state overlaps $\bbrakett{\Psi_{\sigma}}{\Phi_{\rho}}$, where $\kett{\Psi_{\sigma}}$ and $\kett{\Phi_{\rho}}$ are purifications of $\rho$ and $\sigma$ respectively. Using the above definition of $\sigma$, $R_1$ can be written as
\begin{equation}
	R_1(x,y) = \tr[\sqrt{\sigma} \sqrt{\rho}] = \bbrakett{\sqrt{\sigma}}{\sqrt{\rho}} \leq \max_{\kett{\Psi_{\sigma}}, \kett{\Phi_{\rho}}} \bbrakett{\Psi_{\sigma}}{\Phi_{\rho}} = F(x,y) .
\end{equation}
In words, since $\kett{\sigma}$ and $\kett{\rho}$ are themselves purifications of $\sigma$ and $\rho$, their overlap is necessarily no larger than the fidelity between the two states.

To achieve a lower bound, note first that $\sqrt{\sigma} \sqrt{\rho}$ can be written via polar decomposition as $U \sqrt{ \sqrt{\rho} \sigma \sqrt{\rho}}$ for some unitary $U$. This allows for the fidelity to be written as
\begin{equation}
	\begin{split}
		F(x,y) &= \tr[U^{\dag} \sqrt{\sigma} \sqrt{\rho}] = \tr[ (\sigma^{1/4} U \rho^{1/4})^{\dag} (\sigma^{1/4} \rho^{1/4}) ] \\
		&\leq \sqrt{ \tr[(\sigma^{1/4} U \rho^{1/4})^{\dag} (\sigma^{1/4} U \rho^{1/4})] \tr[ (\sigma^{1/4} \rho^{1/4})^{\dag} (\sigma^{1/4} \rho^{1/4}) ] } ,
	\end{split}
\end{equation}
where we have used the cyclicity of the trace and the Cauchy-Schwartz inequality for the Hilbert-Schmidt inner product, $\tr[A^{\dag} B] \leq \sqrt{\tr[A^{\dag} A] \tr[B^{\dag} B] }$. The second term inside the square root is simply $R_1$, while the first term can be upper-bounded using the Cauchy-Schwartz inequality once again:
\begin{equation}
	\tr[(\sigma^{1/4} U \rho^{1/4})^{\dag} (\sigma^{1/4} U \rho^{1/4})] = \tr[ \sqrt{\rho} U^{\dag} \sqrt{\sigma} U ] \leq \sqrt{ \tr[\rho] \tr[U^{\dag} \sigma U] } = 1 .
\end{equation}
We thus arrive at the inequality $F(x,y) \leq \sqrt{R_1(x,y)}$. Altogether, we obtain the following upper and lower bounds on the R\'enyi-1 correlator:

\begin{equation}
\label{eq:fidelity_R1_inequality}
	[F(x,y)]^2 \leq R_1(x,y) \leq F(x,y) ,
\end{equation}
valid whenever $O_x O^{\dag}_y$ is unitary\footnote{For a more general bound which does not assume that $O_x O^{\dag}_y$ is unitary, see Ref.~\cite{liu2024diagnosingstrongtoweaksymmetrybreaking}.}. As a sanity check, note that $0 \leq F(x,y) \leq 1$ and $0 \leq R_1(x,y) \leq 1$, as follows directly from their definitions.

\section{Strong-To-Weak Spontaneous Symmetry-Breaking in the Decohered Quantum Ising Model}
\label{sec:decohered_ising}
In this Appendix we review the decohered quantum Ising model, a paradigmatic example of SWSSB \cite{lee2023quantumcriticality,sala2024spontaneousstrongsymmetrybreaking,lessa2024strongtoweakspontaneoussymmetrybreaking}. Our purpose here is to demonstrate how the R\'enyi-1 correlator \eqref{eq:renyi_1} and fidelity correlator \eqref{eq:supp_fidelity} [as well as the other observables \eqref{eq:rel_entropy}, \eqref{eq:tr_dist}, and \eqref{eq:supp_renyi_2}] can all be efficiently mapped to correlation functions in an effective statistical mechanics model using the formalism of Appendix \ref{subsec:stab_pauli}, \textit{without} the replica trick.

\begin{figure}
	\centering
	\begin{tikzpicture}
		\node at (-2.75, 5.5) {\large (a)};	
		\foreach \x in {0, 1, 2, 3, 4, 5} {
			\foreach \y in {0, 1, 2, 3, 4, 5} {
				\node at (\x-2, \y) {$\ket{+}$};
			}
		}
		\foreach \x in {0, 1, 2, 3, 4} {
			\foreach \y in {0, 1, 2, 3, 4, 5} {
				\draw[line width = 1.0pt, red, line cap = round] (\x - 2 + 0.25, \y) -- (\x - 2 + 0.75, \y);
			}
		}
		\foreach \x in {0, 1, 2, 3, 4, 5} {
			\foreach \y in {0, 1, 2, 3, 4} {
				\draw[line width = 1.0pt, red, line cap = round] (\x - 2, \y + 0.25) -- (\x - 2, \y + 0.75);
			}
		}
		\draw[line width = 1.0pt, black, ->] (0.5 - 2, 5.1) to[out=135, in=200] (0.85 - 2, 5.5);
		\node at (3.25 - 2, 5.5) {$\mathcal{E}^p_{ij}(\rho) = (1-p) \rho + p Z_i Z_j \rho Z_i Z_j$};

		\node at (5.75, 5.5) {\large (b)};	
		\foreach \x in {6.5, 7.5, 8.5, 9.5, 10.5, 11.5} {
			\foreach \y in {0, 1, 2, 3, 4, 5} {
				\node at (\x, \y) {$\ket{+}$};
			}
		}
		\foreach \x in {6.5, 7.5, 8.5, 9.5, 10.5} {
			\foreach \y in {0, 1, 2, 3, 4, 5} {
				\draw[line width = 1.0pt, gray] (\x + 0.25, \y) -- (\x + 0.75, \y);
			}
		}
		\foreach \x in {6.5, 7.5, 8.5, 9.5, 10.5, 11.5} {
			\foreach \y in {0, 1, 2, 3, 4} {
				\draw[line width = 1.0pt, gray] (\x, \y + 0.25) -- (\x, \y + 0.75);
			}
		}

		\draw[blue, line width = 3pt, line cap = round] (7.8, 1) -- (8.2, 1);
		\draw[blue, line width = 3pt, line cap = round] (8.5, 1.3) -- (8.5, 1.7);
		\draw[green, line width = 3pt, line cap = round] (8.5, 2.3) -- (8.5, 2.7);

		\draw[green, line width = 3pt, line cap = round] (9.8, 2) -- (10.2, 2);
		\draw[green, line width = 3pt, line cap = round] (10.5, 1.3) -- (10.5, 1.7);

		\draw[blue, line width = 3pt, line cap = round] (9.8, 4) -- (10.2, 4);

		\filldraw[white, fill=white] (7.25, 0.75) rectangle (7.75, 1.25);
		\filldraw[white, fill=white] (8.25, 2.75) rectangle (8.75, 3.25);
		\filldraw[white, fill=white] (9.25, 3.75) rectangle (9.75, 4.25);
		\filldraw[white, fill=white] (10.25, 3.75) rectangle (10.75, 4.25);
		\filldraw[white, fill=white] (9.25, 1.75) rectangle (9.75, 2.25);
		\filldraw[white, fill=white] (10.25, 0.75) rectangle (10.75, 1.25);

		\node[red] at (7.5, 1) {$\ket{-}$};
		\node[red] at (8.5, 3) {$\ket{-}$};
		\node[red] at (9.5, 4) {$\ket{-}$};
		\node[red] at (9.5, 2) {$\ket{-}$};
		\node[red] at (10.5, 4) {$\ket{-}$};
		\node[red] at (10.5, 1) {$\ket{-}$};

		\draw[green, line width = 3pt, line cap = round] (7.8, 3) -- (8.2, 3);
		\draw[blue, line width = 3pt, line cap = round] (8.5, 2.7) -- (8.5, 2.625);
		\draw[blue, line width = 3pt, line cap = round] (8.5, 2.375) -- (8.5, 2.3);
		\draw[green, line width = 3pt, line cap = round] (7.5, 2.3) -- (7.5, 2.7);
		\draw[green, line width = 3pt, line cap = round] (7.8, 2) -- (8.2, 2);
		\draw[green, line width = 3pt, line cap = round] (9.5, 0.3) -- (9.5, 0.7);
		\draw[green, line width = 3pt, line cap = round] (9.5, 1.3) -- (9.5, 1.7);
		\draw[green, line width = 3pt, line cap = round] (9.8, 0) -- (10.2, 0);
		\draw[green, line width = 3pt, line cap = round] (10.5, 0.3) -- (10.5, 0.7);
		\draw[blue, line width = 3pt, line cap = round] (10.5, 1.3) -- (10.5, 1.375);
		\draw[blue, line width = 3pt, line cap = round] (10.5, 1.7) -- (10.5, 1.625);
		\draw[blue, line width = 3pt, line cap = round] (9.8, 2) -- (9.875, 2);
		\draw[blue, line width = 3pt, line cap = round] (10.2, 2) -- (10.125, 2);

		\draw[green, line width = 3pt, line cap = round] (7.8, 5) -- (8.2, 5);
		\draw[green, line width = 3pt, line cap = round] (7.5, 4.3) -- (7.5, 4.7);
		\draw[green, line width = 3pt, line cap = round] (8.5, 4.3) -- (8.5, 4.7);
		\draw[green, line width = 3pt, line cap = round] (7.8, 4) -- (8.2, 4);

		\node at (12.5, 3) {$\ell:$};
		\node at (12.85, 2.5) {$v = \partial \ell:$};
		\node at (12.5, 2) {$C:$};

		\draw[blue, line width = 3pt, line cap = round] (13, 3) -- (13.5, 3);
		\node[red] at (13.8, 2.5) {$\ket{-}$};
		\draw[green, line width = 3pt, line cap = round] (13, 2) -- (13.5, 2);

		\draw[black, line width = 2pt, ->] (2, -1.5) -- (7.5, -1.5);
		\draw[black, line width = 1pt] (2, -1.65) -- (2, -1.35);
		\node at (2, -1.85) {$0$};
		\node at (7.75, -1.5) {\large $p$};
		\draw[black, line width = 1pt] (7, -1.65) -- (7, -1.35);
		\node at (7, -1.85) {$1/2$};
		\node[red] at (3.5, -1.5) {\large $\bigstar$};
		\node[green] at (5, -1.5) {\large $\bigstar$};
		\node at (3.5, -1.85) {$p_c$};
		\node at (5, -1.85) {$p_{c2}$};

		\node at (1.5, -0.8) {\large (c)};

		\draw[line width = 1.0pt, decorate, decoration = {brace, amplitude = 5pt, mirror, raise=2pt}] (5, -2) -- (7, -2) node[midway, yshift = -15pt] {$R_2(x,y) \sim \mathcal{O}(1)$};
		\draw[line width = 1.0pt, decorate, decoration = {brace, amplitude = 5pt, raise=-5pt}] (3.5, -1) -- (7, -1) node[midway, yshift = 8pt] {$R_1(x,y) = F(x,y) \sim \mathcal{O}(1)$};






	\end{tikzpicture}
	\caption{Schematic depiction of the decohered quantum Ising model and its statistical physics mapping. (a) Starting from an array of qubits in a $d$-dimensional square lattice, each initialized in the state $\ket{+}$, we apply the dephasing channel $\mathcal{E}_{ij}^p$ to each nearest-neighbor pair $\expval{ij}$ of qubits. (b) As in Eq.~\eqref{eq:rho_p_defn}, the resulting density matrix $\rho_p$ can be unraveled into a sum over ``error chains'' $\ell$, corresponding to a collection of links in the lattice (blue). Each such error chain arises with probability $p_{\ell} = (1-p)^{dN-\abs{\ell}} p^{\abs{\ell}}$, and results in a set of phase flips at the collection of sites $v = \partial \ell$ (red). The total probability of achieving the ``syndrome'' $v$ is computed by summing over all error chains with boundary $v$; each such error chain can be achieved by adding a closed loop $C$ (green) to an initial representative error $\ell$. (c) Phase diagram for the various observables \eqref{eq:renyi_1}, \eqref{eq:supp_fidelity}, and \eqref{eq:supp_renyi_2} in the state $\rho_p$, as a function of $p$. The R\'enyi-1 and fidelity correlators exhibit LRO above $p_c$, while the R\'enyi-2 correlator exhibits LRO only above $p_{c2}$. As shown in Ref.~\cite{lessa2024strongtoweakspontaneoussymmetrybreaking}, all states with $p > p_c$ are two-way connected by short-depth symmetric channels, while a stability theorem forbids the two-way connectivity between states with $p < p_c$ and $p > p_c$. Note that for $d = 1$, $p_c = p_{c2} = 1/2$.}
	\label{fig:decohered_Ising}
\end{figure}
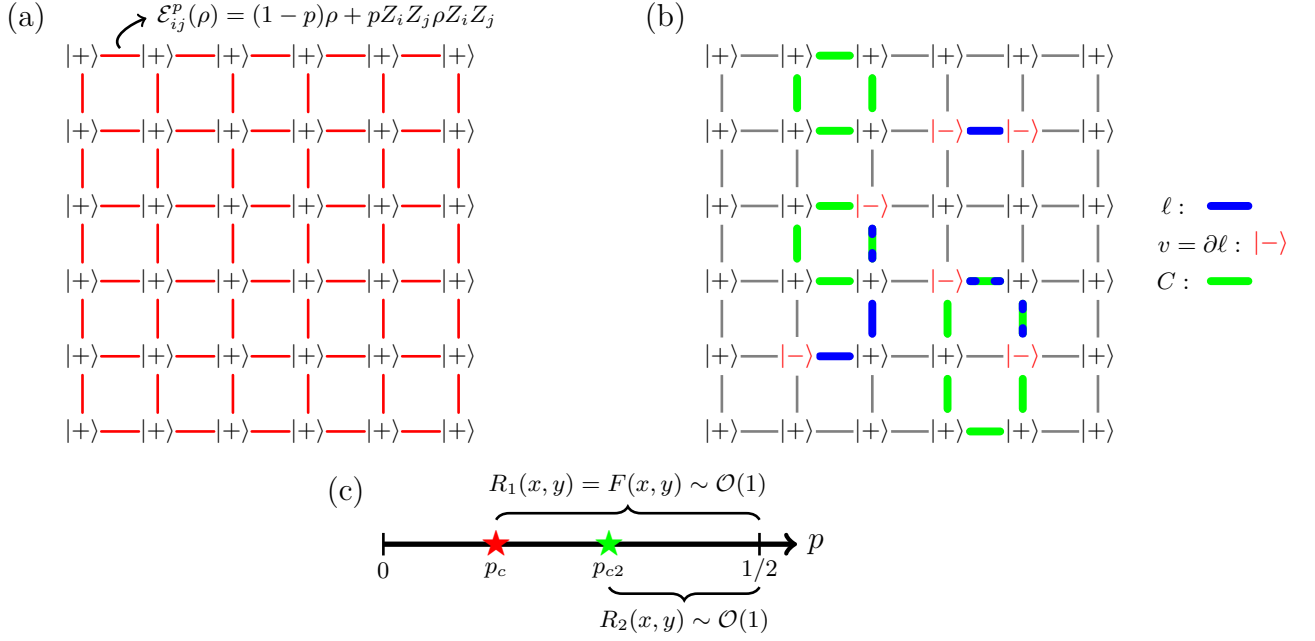

We consider $N$ qubits arranged in a $d$-dimensional square lattice, with periodic boundary conditions for concreteness. The system is initialized in the pure $\mathbb{Z}_2$-symmetric product state $\rho_0 = \dyad{+}^{\otimes N}$, where $\ket{+} \equiv \frac{1}{\sqrt{2}} (\ket{0} + \ket{1})$ is the +1 eigenstate of the Pauli-$X$ operator. We then subject each nearest-neighbor pair of qubits to a $ZZ$ dephasing channel\footnote{Note that in dimension $d = 2$, this problem is exactly the Wegner dual of the toric code decohered by bit-flip errors. The duality is identified by the operator replacements $Z_i Z_j \to X_{ab}$ and $X_j \to B_p$, where $ab$ is a nearest-neighbor pair of sites on the dual lattice bisected by $ij$, and $B_p$ is the toric code plaquette operator at the plaquette $p$ of the dual lattice centered on the site $j$ of the direct lattice. It should therefore be unsurprising that the SWSSB transition in this setting is described by a random-bond Ising model on the Nishimori line.} of strength $p$ [see Fig.~\ref{fig:decohered_Ising}(a)]:
\begin{equation}
	\mathcal{E}^p = \prod_{\expval{ij}} \mathcal{E}^p_{ij}, \quad \mathcal{E}^p_{ij}(\rho) = (1-p) \rho + p Z_i Z_j \rho Z_i Z_j .
\end{equation}
This quantum channel is strongly $\mathbb{Z}_2$-symmetric, in the sense that each Kraus operator commutes with the parity operator $\Pi = \prod_j X_j$. As a result, the decohered state $\rho_p \define \mathcal{E}^p(\rho_0)$ remains strongly symmetric for all values of $p$. Nevertheless, we shall show that $\rho_p$ \textit{spontaneously} breaks this strong symmetry down to a residual weak symmetry for sufficiently large $p$. In $d = 1$ dimension, SWSSB only occurs exactly at the maximal value\footnote{For $p > 1/2$, we can rewrite $\mathcal{E}^p_{ij}(\rho) = \mathcal{E}^{1-p}_{ij}(Z_i Z_j \rho Z_i Z_j)$ as a unitary rotation, followed by the same channel at strength $1-p$. The total channel $\mathcal{E}^p$ does not even require this unitary, i.e., $\mathcal{E}^p = \mathcal{E}^{1-p}$.} $p = 1/2$, while for $d \geq 2$ there is a stable SWSSB phase.


In the language of Appendix \ref{subsec:stab_pauli}, we can think of each error $e$ in the channel \eqref{eq:pauli_channel} with nonzero probability $p_e$ as specified by collection of links $\ell$ in the lattice (i.e., a `1-chain'), visualized as a graph of open strings [see Fig.~\ref{fig:decohered_Ising}(b)]. Each nearest-neighbor link $\expval{ij}$ is included in the graph with probability $p$, and excluded with probability $1-p$. Phase-flip errors $Z_j$ are applied at the endpoints of each included link, and since such errors cancel in even multiples, a given graph $\ell$ applies phase flips only at the collection of its endpoints $v = \partial \ell$ (i.e., a `0-chain'). Thus, each equivalence class $s$ of Appendix \ref{subsec:stab_pauli} is labeled in this setting by the collection $v$ of vertices on which the error acts nontrivially. The general formulae~\eqref{eq:pauli_channel} and~\eqref{eq:pauli_channel_reraveling} are therefore specialized to the present context as follows:
\begin{equation}
\label{eq:rho_p_defn}
	\begin{split}
		\rho_p &\define \mathcal{E}^p(\rho_0) = \sum_{\ell} p_{\ell} \dyad{\partial \ell}, \quad p_{\ell} \define (1-p)^{dN-\abs{\ell}} p^{\abs{\ell}}, \quad \ket{\partial \ell} \define \prod_{j \in \partial \ell} Z_j \ket{+}^{\otimes N} , \\
		&= \sum_{v} P_{v} \dyad{v}, \quad \quad \quad \quad \quad \quad P_{v} \define \sum_{\ell : \partial \ell = v} p_{\ell}, \quad \quad \quad \quad \quad \ket{v} \define \prod_{j \in v} Z_j \ket{+}^{\otimes N} .
	\end{split}
\end{equation}
In the above, each state $\ket{v}$ is simply a $\mathbb{Z}_2$-even product state in the $X$ basis, and $P_{v}$ is the total probability of achieving the state $\ket{v}$ by a sequence of phase-flip errors. 

As a first observation, note that $\bra{v} Z_x Z_y \ket{v} = 0$ for any $x \neq y$, and so ordinary correlation functions of the form $\tr[Z_x Z_y \rho_p]$ trivially have zero correlation length at all values of $p$. On the other hand, correlation functions such as the R\'enyi-1 and fidelity correlators can exhibit nontrivial behavior as a function of $p$ [see Fig.~\ref{fig:decohered_Ising}(c)]. Specifically, the general results from Appendix \ref{subsec:stab_pauli} give
\begin{equation}
\label{eq:R1_F_Ising}
	R_1(x,y) = F(x,y) = \sum_{\ell} p_{\ell} \sqrt{ \frac{P_{\partial \ell \oplus \qty{xy}}}{P_{\partial \ell}} } ,
\end{equation}
where $\oplus$ denotes the mod-two union of the sets of vertices $\partial \ell$ and $\qty{xy}$, i.e., $\partial \ell \oplus \qty{xy} = \partial \ell \cup \qty{xy} - \partial \ell \cap \qty{xy}$. The form of the other correlators can be written down similarly; we focus on $R_1$ and $F$ for purposes of clarity. The probability $P_v$ for fixed $v$ can be computed by starting with a fixed representative error $\ell_v$ which achieves this state (i.e., $v = \partial \ell_v$), and then summing over all other errors $\ell = \ell_v \oplus C$ which differ from the open string $\ell_v$ by a closed loop $C$ [see Fig.~\ref{fig:decohered_Ising}(b)]:
\begin{equation}
\label{eq:P_v}
	P_v = (1-p)^{dN} \sum_{C : \partial C = 0} \qty( \frac{p}{1-p} )^{\abs{\ell_v \oplus C}} ,
\end{equation}
where $\ell_v \oplus C$ denotes the mod-two union of $\ell_v$ and $C$, giving another open string with the same endpoints. Clearly, the sum on the right-hand side is independent of the choice of representative $\ell_v$. We can now proceed to represent $P_v$ as the partition function of a classical statistical physics model in two distinct ways:
\begin{enumerate}
	\item The first approach is to regard the sum over closed loops as a \textit{high-temperature expansion}. This approach has the benefit of treating the models in each spatial dimension $d$ on exactly the same footing: as we shall see, $P_v$ is the partition function of a bond-disordered Ising model in $d$ dimensions (more accurately, a disordered $O(1)$ loop model), and the correlation functions $R_1(x,y)$ and $F(x,y)$ are related to two-point correlation functions averaged over quenched disorder. A potential downside of this approach is that the individual Boltzmann weights of $P_v$ are not all positive in this representation, and therefore not of the form $e^{-\beta H}$ for some classical Hamiltonian $H$.

	\item The second approach is to treat the sum over closed loops as a \textit{low-temperature expansion}. For general dimensions $d \geq 2$, the resulting statistical mechanics model is a disordered version of Wegner's model\footnote{Recall that Wegner's model $M_{d,k}$ is a classical $d$-dimensional lattice spin model with Ising spins on unit $(k-1)$-cells of a hypercubic lattice, with interaction terms in the Hamiltonian taken as the product of all spins belonging to a unit $k$-cell; 0-cells denote vertices, 1-cells denote links, 2-cells denote plaquettes, and so on. In other words, $M_{d,1}$ for $d \geq 1$ is the ferromagnetic Ising model in $d$ dimensions, with spins on vertices and interactions along links of the lattice; $M_{d,2}$ for $d \geq 2$ is the $\mathbb{Z}_2$ lattice gauge theory in $d$ dimensions, with spins on links and interactions on plaquettes; and so on. Wegner's general result was the duality relation between models $M_{d,k}$ and $M_{d,d-k}$, which includes the Kramers-Wannier self-duality of the Ising model in $d = 2$ dimensions, as well as the duality between the Ising model and the $\mathbb{Z}_2$ lattice gauge theory in $d = 3$ dimensions.} $M_{d,d-1}$ \cite{wegnerDualityGeneralizedIsing1971b}; $d = 2$ reproduces the random-bond Ising model on the Nishimori line, while $d = 3$ gives a random-plaquette $\mathbb{Z}_2$ lattice gauge theory with a similar Nishimori-like condition \cite{dennisTopologicalQuantumMemory2002,wangConfinementHiggsTransitionDisordered2003}. These models have previously been studied numerically in detail, and their critical points are known. In this representation, $R_1$ and $F$ are related to \textit{disorder parameter} correlations, i.e., ratios of partition functions with different disorder realizations. 
\end{enumerate}
 We shall now treat both the low-temperature and high-temperature approaches in turn. The two approaches yield statistical physics models which are dual to each other, providing complementary insight on the SWSSB transition in the density matrix $\rho_p$.

\subsection{High-Temperature Expansion}
\label{subsec:high_temp}
Our first method of representing $P_v$ as a classical statistical physics model is to consider the closed loops $C$ as arising from a high-temperature expansion of an Ising system. In other words, we place a classical Ising spin $\sigma_j = \pm 1$ on each site of the lattice, and represent $P_v$ as the following partition function with quenched disorder:
\begin{equation}
\label{eq:P_v_2}
	P_v = \frac{(1-p)^{dN}}{2^N} \qty(\frac{p}{1-p})^{\abs{\ell_v}} \sum_{\qty{\sigma}} \prod_{\expval{ij}} \qty( 1 + t_{ij} \sigma_i \sigma_j ), \quad t_{ij} = \begin{cases}
		p/(1-p), & \expval{ij} \not\in \ell_v \\
		(1-p)/p, & \expval{ij} \in \ell_v
	\end{cases} .
\end{equation}
A high-temperature expansion of the above partition function sum then yields a sum over graphs of closed loops in the lattice, with a factor $t_{ij}$ for each link $\expval{ij}$ included in a graph. In this representation, independence of the choice of representative $\ell_v$ follows from inserting $1 = \prod_{\expval{ij} \in C} \sigma_i \sigma_j$ for a closed curve $C$ into the Boltzmann weight, which modifies the bonds along the curve $C$ as follows:
\begin{equation}
	\begin{split}
		\prod_{\expval{ij} \in C} (1 + t_{ij} \sigma_i \sigma_j) &= \qty( \prod_{\expval{ij} \in C} \sigma_i \sigma_j ) \qty( \prod_{\expval{ij} \in C} (1 + t_{ij} \sigma_i \sigma_j) ) = \prod_{\expval{ij} \in C} (\sigma_i \sigma_j + t_{ij}) \\
		&= \qty( \prod_{\expval{ij} \in C} t_{ij} ) \qty( \prod_{\expval{ij} \in C} (1 + t_{ij}^{-1} \sigma_i \sigma_j) ) .
	\end{split}
\end{equation}
The factor $\prod_{\expval{ij} \in C} t_{ij}$ then combines with the prefactor $[p/(1-p)]^{\abs{\ell_v}}$ to give $[p/(1-p)]^{\abs{\ell_v \oplus C}}$. By a similar calculation, inserting a factor $\sigma_x \sigma_y = \prod_{\expval{ij} \in \gamma} \sigma_i \sigma_j$ for an \textit{open} string $\gamma$ with endpoints $x$ and $y$ gives the following result for two-point correlations in this model:
\begin{equation}
\label{eq:ising_corr}
	\expval{\sigma_x \sigma_y}_{\ell_v} \define \frac{\sum_{\qty{\sigma}} \sigma_x \sigma_y \prod_{\expval{ij}} (1 + t_{ij} \sigma_i \sigma_j)}{\sum_{\qty{\sigma}} \prod_{\expval{ij}}(1 + t_{ij} \sigma_i \sigma_j)} = \frac{P_{v \oplus \qty{xy}}}{P_v} .
\end{equation}
From Eq.~\eqref{eq:R1_F_Ising}, we immediately obtain the following result for the R\'enyi-1 correlator and the fidelity correlator:
\begin{equation}
	\boxed{ R_1(x,y) = F(x,y) = \sum_{\ell} p_{\ell} \sqrt{\frac{P_{\partial \ell \oplus \qty{xy}}}{P_{\partial \ell}}} = \sum_{\ell} p_{\ell} \sqrt{ \expval{\sigma_x \sigma_y}_{\ell} } . }
\end{equation}
In the final expression, we can regard the average over $\ell$ as a simple quenched disorder average in the effective statistical mechanics model, with the value of each nearest-neighbor coupling $t_{ij}$ sampled independently. Note that the magnitude of $t_{ij}$ is a function of the bond disorder probability $p$, providing a ``Nishimori-like'' condition. As we will see below, the $d = 2$ model is exactly the Kramers-Wannier dual of the random-bond Ising model on the Nishimori line.

The partition function \eqref{eq:P_v_2} is best interpreted as a disordered $O(1)$ loop model on the square lattice; strictly speaking, it is not an Ising model with Boltzmann weights of the form $e^{-\beta H}$, since $t_{ij}$ can be greater than one. Nevertheless, the qualitative phase diagram and the behavior of $R_1$ and $F$ can easily be determined from this mapping. For small $p$ the model is in a paramagnetic phase, and correlations $\expval{\sigma_x \sigma_y}_{\ell}$ decay exponentially in $\abs{x-y}$ for typical realizations of $\ell$. On the other hand, for sufficiently large $p$ and in dimensions $d \geq 2$, the model is expected to exhibit a ferromagnetic phase with long-range order, and $R_1$ and $F$ asymptotically give the disorder-averaged order parameter $\sum_{\ell} p_{\ell} \abs{\expval{\sigma}_{\ell}}$ as $\abs{x-y} \to \infty$. We thus expect a SWSSB transition in $\rho_p$ as a function of $p$ for spatial dimensions $d \geq 2$.

In the case $d = 1$ we can compute $R_1$ and $F$ exactly in the thermodynamic limit $N \to \infty$, allowing for us to verify that $\rho_p$ exhibits SWSSB only at $p = 1/2$. For the correlation function, we have
\begin{equation}
	\expval{\sigma_0 \sigma_r}_{\ell} = \frac{\prod_{j = 1}^{r} t_{j, j+1} + \prod_{j = r + 1}^{N}t_{j, j+1}}{1 + \prod_{j = 1}^N t_{j, j+1}} \stackrel{r \to \infty}{\simeq} \prod_{j = 1}^r t_{j, j+1} \quad (d=1) ,
\end{equation}
where we've kept only the leading term as $r \to \infty$. After taking the square root, the disorder average over each bond variable $\sqrt{t_{j,j+1}}$ can be performed independently. We therefore find the final result
\begin{equation}
	R_1(x,y) = F(x,y) \stackrel{r \to \infty}{\simeq} \qty[ (1 - p) \sqrt{ \frac{p}{1-p} } + p \sqrt{ \frac{1-p}{p} } ]^{\abs{x-y}} = \qty[ 2\sqrt{p(1-p)} ]^{\abs{x-y}} \quad (d=1) ,
\end{equation}
which exhibits exponential decay in $\abs{x-y}$ for all $p < 1/2$. At exactly $p = 1/2$, it is clear that we have perfect long-range order in $\expval{\sigma_x \sigma_y}_{\ell}$ and SWSSB in $\rho_p$: indeed, the partition function \eqref{eq:P_v_2} becomes that of a clean zero-temperature Ising model.

\subsection{Low-Temperature Expansion}
To gain further insight on the models in dimensions $d \geq 2$, it is useful to instead treat the sum over closed loops in Eq.~\eqref{eq:P_v} as a low-temperature expansion. This gives a statistical physics model which is dual to the one described in the previous section. A benefit of this approach is that it yields a proper disordered Hamiltonian model, with positive Boltzmann weights of the form $e^{-\beta H}$; in principle this makes the phases of $\rho_p$ easier to interpret, although the models are somewhat baroque in dimensions above $d > 3$. For this reason, rather than working abstractly in general dimensions $d$, it is useful here to focus on the concrete examples $d = 2,3$ in turn. We shall simply state the general result that the partition function $P_v$ is described in a low-temperature expansion as a disordered version of the Wegner model $M_{d,d-1}$ \cite{wegnerDualityGeneralizedIsing1971b}.

\subsubsection{\texorpdfstring{$d = 2$}{d=2}: Random-Bond Ising Model on the Nishimori Line}
Let us first concentrate on the simpler case $d = 2$. Here we think of the sum over closed loops in Eq.~\eqref{eq:P_v} as a sum over one-dimensional domain walls of Ising variables. If we momentarily neglect non-contractible loops in this sum, then we can write each closed loop $C$ as $\partial R$, the boundary of a two-dimensional set of plaquettes $R$ (i.e., a `2-chain'). We can sum over all such collections of plaquettes by placing an Ising spin $\mu_a$ at the center of each plaquette $a$, with $\mu_a = +1$ ($\mu_a = -1$) denoting the exclusion (inclusion) of the plaquette $a$ in $R$. The boundaries of the region $R$ are then simply given by the bonds $\expval{ab}$ in the dual lattice for which $\mu_a \mu_b = -1$. With this notation in mind, we have the following:
\begin{equation}
	\begin{split}
		\sum_{C : C = \partial R} \qty( \frac{p}{1-p} )^{\abs{\ell_v \oplus C}} &= \frac{1}{2} \sum_{R} \qty( \frac{p}{1-p} )^{\abs{\ell_v \oplus \partial R}} = \frac{1}{2} \sum_{\qty{\mu}} \prod_{\expval{ab}} \qty( \frac{p}{1-p} )^{(1 - \eta_{ab} \mu_a \mu_b)/2} \\
		&= \frac{1}{2} \qty( \frac{p}{1-p} )^{N} \sum_{\qty{\mu}} \exp{ J \sum_{\expval{ab}} \eta_{ab} \mu_a \mu_b },
	\end{split}
\end{equation}
where we have introduced the bond disorder variable $\eta_{ab} = \pm 1$ and the bond strength $J$, defined respectively by
\begin{equation}
\label{eq:lowtemp_bond_disorder}
	\eta_{ab} = \begin{cases}
			+1, & \expval{ab} \not\in \ell_v \\
			-1, & \expval{ab} \in \ell_v
		\end{cases}, \quad e^{-2J} = \frac{p}{1-p} .
\end{equation}
The above is simply the partition function of a random-bond Ising model (RBIM) along the Nishimori line. The expression $\eta_{ab} \mu_a \mu_b = \pm 1$ simply counts whether the direct-lattice link bisected by the dual-lattice link $\expval{ab}$ is included in $\ell_v \oplus \partial R$ an even ($\eta_{ab} \mu_a \mu_b = +1$) or odd ($\eta_{ab} \mu_a \mu_b = -1$) number of times.

To obtain the full expression for $P_v$, we must also allow for topologically nontrivial loops $C$. This is effectively implemented by summing over both periodic and antiperiodic boundary conditions in both spatial directions. More precisely, we define a pair of non-contractible loops $\Gamma_x$ and $\Gamma_y$ which encircle the two cycles of the torus, and perform the sum as follows:
\begin{equation}
	\begin{split}
		\sum_{C: \partial C = 0} \qty( \frac{p}{1-p} )^{\abs{\ell_v \oplus C}} \! \! &= \! \! \sum_{C : C = \partial R} \qty[ \qty( \frac{p}{1-p} )^{\abs{\ell_v \oplus C}} \! \! + \qty( \frac{p}{1-p} )^{\abs{\ell_v \oplus C \oplus \Gamma_x}} \! \! + \qty( \frac{p}{1-p} )^{\abs{\ell_v \oplus C \oplus \Gamma_y}} \! \! + \qty( \frac{p}{1-p} )^{\abs{\ell_v \oplus C \oplus \Gamma_x \oplus \Gamma_y}} ] \\
		&= \frac{1}{2} \qty( \frac{p}{1-p} )^N \sum_{\qty{\mu}} \qty[ e^{J \sum_{\expval{ab}} \eta_{ab} \mu_a \mu_b } + e^{J \sum_{\expval{ab}} \eta^{(x)}_{ab} \mu_a \mu_b } + e^{J \sum_{\expval{ab}} \eta^{(y)}_{ab} \mu_a \mu_b } + e^{J \sum_{\expval{ab}} \eta^{(xy)}_{ab} \mu_a \mu_b } ] \\
		&\equiv \frac{1}{2} \qty( \frac{p}{1-p} )^N \sum_{\qty{\mu}}' e^{J \sum_{\expval{ab}} \eta_{ab} \mu_a \mu_b } .
	\end{split}
\end{equation}
Here $\eta^{(x)}_{ab}$, $\eta^{(y)}_{ab}$, and $\eta^{(xy)}_{ab}$ are bond disorder configurations with additional `fluxes' inserted through the holes of the torus; i.e., they are given by the same expression as $\eta_{ab}$ in Eq.~\eqref{eq:lowtemp_bond_disorder}, but with $\ell_v$ replaced with $\ell_v \oplus \Gamma_x$, $\ell_v \oplus \Gamma_y$, and $\ell_v \oplus \Gamma_x \oplus \Gamma_y$ respectively. Since this technical complication of summing over non-contractible cycles will not matter for any thermodynamic properties, we have hidden it in the final expression, where we use the primed sum to denote an additional sum over periodic and antiperiodic boundary conditions. With these comments in mind, the probabilities $P_v$ in the $d = 2$ dimensional model are given by the following partition function sum:
\begin{equation}
	P_v = \frac{1}{2} \qty[p(1-p)]^N \sum_{\qty{\mu}}' \exp \qty{ J \sum_{\expval{ab}} \eta_{ab} \mu_a \mu_b } .
\end{equation}

The original expression for $P_v = \sum_{\ell : \partial \ell = v} p_{\ell}$ is obtained by performing a low-temperature expansion of the above partition function, i.e., an ordinary sum over domain-wall configurations of the $\mu_a$ spins. If we regard a dual-lattice link $\expval{ab}$ along which $\eta_{ab} \mu_a \mu_b = -1$ as detecting a segment of domain wall along the direct lattice, then the low-temperature expansion of $P_v$ consists of a sum over \textit{open-string} domain walls. The endpoints of these open strings are ``Ising fluxes'', that is, dual-lattice plaquettes $[abcd]$ for which $\eta_{ab} \eta_{bc} \eta_{cd} \eta_{da} = -1$; from Eq.~\eqref{eq:lowtemp_bond_disorder}, these are precisely the set of points $v$ in the direct lattice. Finally, the independence of the choice of representative $\ell_v$ arises from the invariance of $P_v$ under the ``gauge transformation'' $\eta_{ab} \to \tau_a \eta_{ab} \tau_b$ for a set of numbers $\tau_a = \pm 1$, which can be compensated by the change of variables $\mu_a \to \tau_a \mu_a$ in the partition function sum.

From Eq.~\eqref{eq:R1_F_Ising}, the R\'enyi-1 and fidelity correlators are related to the ratio of partition functions $P_{v \oplus \qty{xy}} / P_v$. We found in Eq.~\eqref{eq:ising_corr} that this quantity was given by the two-point correlation function $\expval{\sigma_x \sigma_y}_{\ell_v}$ in terms of the $\sigma_j$ degrees of freedom. In the $\mu_a$ degrees of freedom, this ratio takes the form of a \textit{disorder parameter} correlation function, in which we compute the free energy cost of inserting an extra pair of Ising fluxes at the dual lattice plaquettes $x$ and $y$. Explicitly, we consider a curve $\gamma$ through the direct lattice (i.e., the dual of the dual lattice) with endpoints at sites $x$ and $y$, and flip the sign of the bond disorder variables $\eta_{ab}$ along each dual-lattice bond $\expval{ab}$ bisected by $\gamma$. This is equivalent to modifying the disorder realization $\ell_v$ to $\ell_v \oplus \gamma$. The ratio of the resulting partition function to the original one defines the disorder parameter correlation function:
\begin{equation}
	\frac{P_{v \oplus \qty{xy}}}{P_v} = \frac{\sum_{\qty{\mu}}' \exp \qty{ J \sum_{\expval{ab}} \tilde{\eta}_{ab} \mu_a \mu_b } }{\sum_{\qty{\mu}}' \exp \qty{ J \sum_{\expval{ab}} \eta_{ab} \mu_a \mu_b } }, \quad \tilde{\eta}_{ab} = \begin{cases}
		+ \eta_{ab}, & \eta_{ab} \not\in \gamma \\
		- \eta_{ab}, & \eta_{ab} \in \gamma
	\end{cases} .
\end{equation}
In the large $J$ (small $p$) phase of the RBIM, the spins $\mu_a$ are in a ferromagnetic phase. Modifying the disorder by $\gamma$ effectively inserts an open-string domain wall from $x$ to $y$, whose free energy $\delta F$ grows linearly with the separation $\abs{x-y}$. The above ratio of partition functions can be interpreted as $e^{-\delta F}$, and so $P_{v \oplus \qty{xy}} / P_v$ decays exponentially in $\abs{x-y}$ for typical disorder realizations. On the other hand, in the small $J$ (large $p$) phase, the $\mu_a$ spins are in a paramagnetic phase, and it costs an order-one free energy to insert such a domain wall into the system. Consequently, $P_{v \oplus \qty{xy}} / P_v$ is of order unity in typical realizations, leading to long-range order in the correlators \eqref{eq:R1_F_Ising}. We thus recover the same qualitative phase diagram as we found in Sec.~\ref{subsec:high_temp}: at some critical dephasing strength $p_c$, we expect a phase transition to a SWSSB phase above which the R\'enyi-1 and fidelity correlators become long-range ordered.

The true benefit of the above RBIM representation of $P_v$ is the vast literature available on the critical phenomena of the RBIM on the Nishimori line. For example, we can immediately conclude that the critical value of $p$ above which the SWSSB phase arises is approximately $p_c \approx 0.109$ \cite{honecker2001universality}. 

\subsubsection{\texorpdfstring{$d = 3$}{d=3}: Random-Plaquette \texorpdfstring{$\mathbb{Z}_2$}{Z2} Gauge Theory}
In $d = 3$ dimensions, we interpret the sum over closed loops in Eq.~\eqref{eq:P_v} as a sum over flux tubes in a $\mathbb{Z}_2$ gauge theory. Let us once again neglect non-contractible loops for the moment and regard closed loops $C$ as the boundary of a 2-chain $R$, which should be pictured as a collection of plaquettes in the direct lattice. To enumerate all such choices, we place an Ising spin $\mu_{ab}$ at the center of each nearest-neighbor link $\expval{ab}$ of the dual lattice, which passes directly through a plaquette of the direct lattice. We set $\mu_{ab} = +1$ ($\mu_{ab} = -1$) whenever the plaquette pieced by $\expval{ab}$ is included (excluded) in $R$. The boundaries of $R$ are then found by taking the product of spins $\mu_{ab}$ around a dual-lattice plaquette; this product is $+1$ $(-1)$ whenever the direct-lattice edge which pierces the dual-lattice plaquette belongs to $\partial R$. Thus, we obtain the following representation of the sum over closed loops:
\begin{equation}
	\begin{split}
		\sum_{C : C = \partial R} \qty( \frac{p}{1-p} )^{\abs{\ell_v \oplus C}} &= \frac{1}{2^N} \sum_R \qty( \frac{p}{1-p} )^{\abs{\ell_v \oplus \partial R}} = \frac{1}{2^N} \sum_{\qty{\mu}} \prod_{[abcd]} \qty( \frac{p}{1-p} )^{(1-\eta_{abcd} \mu_{ab} \mu_{bc} \mu_{cd} \mu_{da})/2} \\
		&= \frac{1}{2^N} \qty( \frac{p}{1-p} )^{3N/2} \sum_{\qty{\mu}} \exp{ J \sum_{[abcd]} \eta_{abcd} \mu_{ab} \mu_{bc} \mu_{cd} \mu_{da} } .
	\end{split}
\end{equation}
This is the partition function of the random-plaquette $\mathbb{Z}_2$ gauge theory, introduced in Refs.~\cite{dennisTopologicalQuantumMemory2002,wangConfinementHiggsTransitionDisordered2003}. The coupling constant $J$ satisfies the same Nishimori condition as before; the plaquette disorder variable $\eta_{abcd}$ tracks whether the dual-lattice plaquette $[abcd]$ is pierced by the links $\ell_v$:
\begin{equation}
\label{eq:bond_disorder_3d}
	\eta_{abcd} = \begin{cases}
		+1, & [abcd] \not\in \ell_v \\
		-1, & [abcd] \in \ell_v
	\end{cases} .
\end{equation}

Once again, to obtain $P_v$ we must modify this sum to include non-contractible closed loops $C$; in the present context, this amounts to allowing odd numbers of non-contractible flux tubes to traverse the periodic boundaries of the system. With analogous primed sum notation to the previous section, $P_v$ is now given by
\begin{equation}
	P_v = \frac{1}{2^N} \qty[ p(1-p) ]^{3N/2} \sum_{\qty{\mu}}' \exp{ J \sum_{[abcd]} \eta_{abcd} \mu_{ab} \mu_{bc} \mu_{cd} \mu_{da} } .
\end{equation}

As before, the original form of $P_v$ is recovered by performing a low-temperature expansion of the above partition function sum. If we regard the dual-lattice plaquettes $[abcd]$ for which $\eta_{abcd} \mu_{ab} \mu_{bc} \mu_{cd} \mu_{da} = -1$ as containing $\mathbb{Z}_2$ gauge flux, then the low-temperature expansion of $P_v$ consists of a sum over open-string flux tubes. The endpoints of these flux tubes are ``magnetic monopoles'', i.e., cubes $\mathcal{C}$ in the dual lattice (centered on sites of the direct lattice) for which $\prod_{[abcd] \in \mathcal{C}} \eta_{abcd} = -1$. From Eq.~\eqref{eq:bond_disorder_3d}, these monopoles lie at precisely the sites $v$ of the error syndromes in the direct lattice. Finally, the independence of the choice of representative $\ell_v$ arises from the invariance of $P_v$ under the ``higher-form'' gauge transformation $\eta_{abcd} \to \tau_{ab} \tau_{bc} \tau_{cd} \tau_{da} \eta_{abcd}$ for some set of numbers $\tau_{ab} = \pm 1$ on each link of the dual lattice; this transformation can be compensated by the change of variables $\mu_{ab} \to \tau_{ab} \mu_{ab}$ in the partition function sum.

Just as in the $d = 2$ case, the correlation function $\expval{\sigma_x \sigma_y}_{\ell} = P_{\partial \ell + \qty{xy}} / P_{\partial \ell}$ maps to a disorder parameter correlation function in the above dual representation. In the three-dimensional case, this corresponds to the free energy cost of inserting two extra magnetic monopoles at the cubes centered on $x$ and $y$. For large $J$ (small $p$) the lattice gauge theory is in a deconfined phase, and magnetic flux is expelled; consequently, the flux tube costs a free energy linear in its length, and $\expval{\sigma_x \sigma_y}_{\ell}$ decays exponentially in $\abs{x-y}$ for typical realizations of $\ell$. On the other hand, for small $J$ (large $p$) the lattice is in a confined phase, and flux tubes are proliferated. Inserting two far-separated extra monopoles therefore costs an order-one free energy, and $\expval{\sigma_x \sigma_y}_{\ell}$ is typically of order-one as $\abs{x - y} \to \infty$. We therefore once again recover the qualitative phase diagram predicted in Sec.~\ref{subsec:high_temp}: at some critical dephasing strength $p_c$, we expect a SWSSB phase transition above which the R\'enyi-1 and fidelity correlators become long-range ordered.

Once again, the critical phenomena of the $\mathbb{Z}_2$ random-plaquette gauge theory has been studied extensively in connection to error thresholds in surface codes \cite{dennisTopologicalQuantumMemory2002,wangConfinementHiggsTransitionDisordered2003,ohnoPhaseStructureRandomplaquette2004}. For example, numerical simulations indicate a critical error rate of $p_c \approx 0.033$ \cite{ohnoPhaseStructureRandomplaquette2004}.

\subsection{R\'enyi-2 Correlations and Annealed Averages}
We finish by commenting on the behavior of the R\'enyi-2 correlation function $R_2(x,y)$, defined in Eq.~\eqref{eq:supp_renyi_2}. From the general discussion of Sec.~\ref{subsec:stab_pauli}, we expect that $R_2(x,y)$ will be given by an \textit{annealed} average correlation function rather than a quenched average; correspondingly, it exhibits the critical phenomena of a clean $d$-dimensional Ising model. 

Although $R_2$ can again be calculated using the formalism of Sec.~\ref{subsec:stab_pauli}, a simpler approach is to expand $\rho_0$ and $\rho_p$ in the Pauli-$Z$ basis. Using $\ket{+} = \frac{1}{\sqrt{2}} (\ket{0} + \ket{1})$, we have
\begin{equation}
	\begin{split}
		\rho_0 = \frac{1}{2^N} \sum_{\qty{\sigma, \sigma'}} \ket{\qty{\sigma}} \bra{\qty{\sigma'}}, \quad \rho_p &= \frac{1}{2^N} \sum_{\qty{\sigma, \sigma'}} \prod_{\expval{ij}} \qty[ (1-p) + p \sigma_i \sigma_i' \sigma_j \sigma_j' ] \ket{\qty{\sigma}} \bra{\qty{\sigma'} } \\
		&= \frac{1}{2^N} \frac{(1-p)^{dN}}{\cosh^{dN} (\beta/2)} \sum_{\qty{\sigma, \sigma'}} \exp \qty{ \frac{\beta}{2} \sum_{\expval{ij}} \sigma_i \sigma_i' \sigma_j \sigma_j' } \ket{\qty{\sigma}} \bra{\qty{\sigma'}} ,
	\end{split}
\end{equation}
where $\ket{\qty{\sigma}} = \ket{\sigma_1 \ldots \sigma_N}$ is a computational basis state\footnote{Note the abuse of notation: $\sigma = \pm 1$, but $\ket{\sigma = +1}$ corresponds to the state $\ket{0}$ and $\ket{\sigma = -1}$ corresponds to the state $\ket{1}$. One should not get confused, for example, in thinking that $\sigma_j$ can take the value zero.}, and $\beta$ is defined via the relation $\tanh \beta/2 = p / (1-p)$. the purity is therefore given by
\begin{equation}
	\tr \rho_p^2 = \qty[ \frac{1}{2^N} \frac{(1-p)^{dN}}{\cosh^{dN}(\beta / 2)} ]^2 \sum_{\qty{\sigma, \sigma'}} \exp{ \beta \sum_{\expval{ij}} \sigma_i \sigma'_i \sigma_j \sigma'_j } = \qty[ \frac{1}{2^N} \frac{(1-p)^{dN}}{\cosh^{dN}(\beta / 2)} ]^2 2^N \sum_{\qty{\sigma}} \exp{ \beta \sum_{\expval{ij}} \sigma_i \sigma_j } .
\end{equation}
In the final expression, we have performed the change of summation variables $\sigma_j \sigma_j' \to \sigma_j$, after which the $\qty{\sigma'}$ summation drops out completely. Thus, the purity $\tr \rho_p^2$ is proportional to the partition function of a $d$-dimensional Ising model at inverse temperature\footnote{As a sanity check, the limit $p \to 0$ corresponds to the infinite temperature limit $\beta \to 0$, which gives $\tr \rho_p^2 = 1$ as desired. On the other hand, the limit $p \to 1/2$ corresponds to $\beta \to \infty$, which gives the purity $\tr \rho_p^2 = 2^{-(N-1)}$ as expected.} $\beta$. Meanwhile, the numerator of $R_2$ is obtained by inserting $Z_x Z_y$ on both sides of one factor of $\rho_p$, which gives the same expression as above with $\sigma_x \sigma_x' \sigma_y \sigma_y'$ inserted.  After the same change of summation variables, this becomes simply $\sigma_x \sigma_y$. We therefore find that $R_2$ is given simply by the correlation function of a $d$-dimensional Ising model:
\begin{equation}
	R_2(x,y) = \frac{\tr[Z_x Z_y \rho_p Z_x Z_y \rho_p]}{\tr \rho_p^2} = \frac{\sum_{\qty{\sigma}} \sigma_x \sigma_y e^{\beta \sum_{\expval{ij}} \sigma_i \sigma_j}}{\sum_{\qty{\sigma}} e^{\beta \sum_{\expval{ij}} \sigma_i \sigma_j}} .
\end{equation}
In two dimensions, $R_2$ exhibits a phase transition at $\beta_{c2} = \frac{1}{2} \log(1 + \sqrt{2})$, corresponding to $p_{c2} \approx 0.178$. Notably, this value is larger than the RBIM critical point which controls the R\'enyi-1 and fidelity correlators, implying a contiguous region of SWSSB without long-range order in the R\'enyi-2 correlator. Similarly, the phase transition occurs in three dimensions at $\beta_{c2} \approx 0.221$ \cite{talapovMagnetization3DIsing1996}, corresponding to $p_{c2} \approx 0.099$. This is once again larger than the critical point of the random-plaquette gauge theory.

\section{Strong-To-Weak Spontaneous Symmetry Breaking in the One-Dimensional Transverse-Field Ising Model at Nonzero Temperatures}
In this section, we demonstrate that the thermal Gibbs state of the one-dimensional transverse-field Ising model (TFIM) exhibits SWSSB at any nonzero temperature. As briefly mentioned in the main text (and discussed in more detail in Ref.~\cite{lessa2024strongtoweakspontaneoussymmetrybreaking}), SWSSB is expected to be a generic feature of the symmetry-restored phase of thermal Gibbs states. This is easily verified in the case of commuting projector Hamiltonians. The exact solvability of the one-dimensional TFIM allows us to additionally verify this feature in a simple non-commuting setting.

We consider $N$ qubits in one spatial dimension with periodic boundary conditions. The Hamiltonian is
\begin{equation}
	H = -J \sum_{j = 1}^N \Big\{ Z_j Z_{j+1} + g X_j \Big\} ,
\end{equation}
with $Z_{N+1} \equiv Z_1$. This model exhibits a $\mathbb{Z}_2$ parity symmetry generated by $\Pi = \prod_{j = 1}^N X_j$. Our focus is on observables in the parity-even sector of the thermal Gibbs state, with the density matrix
\begin{equation}
\label{eq:rho_beta_evenparity}
	\rho_{\beta} = \frac{P_{\Pi} e^{-\beta H}}{\mathcal{Z}_{\beta}}, \quad \mathcal{Z}_{\beta} = \tr[ P_{\Pi} e^{-\beta H}], \quad P_{\Pi} = \frac{1 + \Pi}{2} .
\end{equation}

The TFIM is exactly solvable via the Jordan-Wigner transformation. We define the Majorana fermion operators
\begin{equation}
	\gamma_{2j-1} = \qty[ \prod_{i = 1}^{j-1} X_i ] Z_j, \quad \gamma_{2j} = \qty[ \prod_{i = 1}^{j - 1} X_i ] Y_j ,
\end{equation}
which satisfy the anticommutation relations $\acomm{\gamma_i}{\gamma_j} = 2\delta_{ij}$, as well as the useful identities $X_j = i \gamma_{2j-1} \gamma_{2j}$ and $Z_j Z_{j+1} = i \gamma_{2j} \gamma_{2j+1}$. In terms of the Majoranas, $H$ is written as
\begin{equation}
	H = -iJg \sum_{j = 1}^N \gamma_{2j-1} \gamma_{2j} - iJ \sum_{j = 1}^{N-1} \gamma_{2j} \gamma_{2j+1} + i J \Pi \gamma_{2N} \gamma_1 .
\end{equation}
In other words, $H$ is a model of free Majorana fermions with nearest-neighbor couplings of alternating strength. From the last term, we see that the Majorana chain has antiperiodic (periodic) boundary conditions in the parity-even (parity-odd) sector. Since $\rho_{\beta}$ is projected into the parity-even sector by design, we may simply set $\Pi = 1$ and use antiperiodic boundary conditions. Our Hamiltonian is therefore a simple quadratic Majorana fermion Hamiltonian, of the general form
\begin{equation}
	H = \frac{i}{4} \sum_{i,j = 1}^{2N} \gamma_i A_{ij} \gamma_j ,
\end{equation}
where $A$ is a $2N \times 2N$ real antisymmetric matrix. Arbitrary correlation functions of $H$ can then be efficiently computed numerically for large values of $N$; see Appendix A of Ref.~\cite{weinsteinNonlocalityEntanglementMeasured2023a} for a review.

Our goal is to compute the following R\'enyi-1 correlator:
\begin{equation}
	\begin{split}
		R_1(x,y) &= \tr[ Z_x Z_y \sqrt{\rho_{\beta}} Z_x Z_y \sqrt{\rho_{\beta}} ] \\
		&= \frac{1}{\mathcal{Z}_{\beta}} \tr[ P_{\Pi} Z_x Z_y e^{-\beta H/2} Z_x Z_y e^{-\beta H/2} ] \\
		&= \frac{1}{\mathcal{Z}_{\beta}} \tr \qty[ P_{\Pi} \qty( e^{\beta H/2} Z_x Z_y e^{-\beta H/2} ) Z_x Z_y e^{-\beta H} ] .
	\end{split}
\end{equation}
In other words, $R_1$ is simply an imaginary time-ordered four-point correlation function, with two of the operators evolved to imaginary time $\tau = \beta / 2$. Before moving onto the general computation, let us compute $R_1(x,y)$ in two extreme limits: deep in the ferromagnetic regime with $g \to 0$, and deep in the paramagnetic regime $J \to 0$, $Jg \equiv h = \text{const}$. In the former limit, $Z_x Z_y$ commutes with $\sqrt{\rho_{\beta}}$, and so we trivially have $R_1(x,y) = 1$. In the latter limit, we can use the result of Sec.~\ref{subsec:gibbs} to immediately obtain $R_1(x,y) = \sech^2(\beta h)$. Thus, we find in both regimes that SWSSB arises at at any nonzero temperature.

\begin{figure}[t]
	\centering
	\includegraphics[width = 0.45\textwidth]{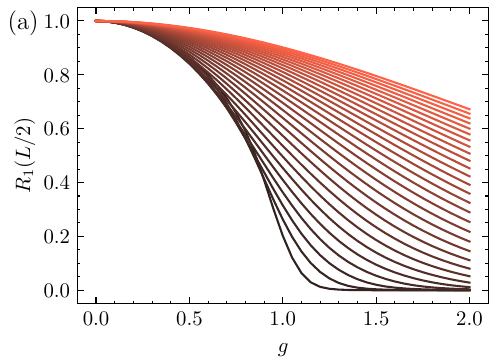}
	\includegraphics[width = 0.45\textwidth]{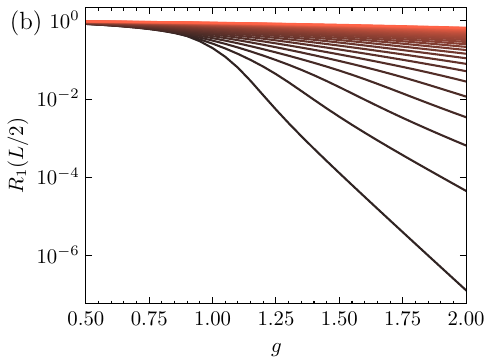}
	\caption{Numerical results for the R\'enyi-1 correlator $R_1(x,y)$ for $\abs{x-y} = L/2$ in the finite-temperature Gibbs state of the transverse-field Ising model on $L$ sites; here $L = 128$. Note that the density matrix is restricted to the parity-even sector to achieve a strongly-symmetric density matrix, as in Eq.~\eqref{eq:rho_beta_evenparity}. Each curve corresponds to a different temperature, in the range $[0.15J, 2.95J]$ in intervals of $0.1J$. Darker colors denote lower temperatures. (a) For very low temperatures, $R_1(x,y)$ exhibits crossover behavior in the vicinity of the zero-temperature transition point $g = 1$; in the limit $T \to 0$, $R_1(L/2)$ would exactly mimic the behavior of $\expval{Z}^4$. As the temperature is raised, the crossover is smoothed further. (b) The same plot as in (a), with the vertical axis on a logarithmic scale. We see that $R_1(L/2)$ decays exponentially in $g$ with a slope set by the inverse temperature, as expected from the $J \to 0$ analytical solution for $R_1(x,y)$.}
	\label{fig:ising_plots}
\end{figure}

We now move onto an efficient numerical free fermion method to compute $R_1(x,y)$ for general values of $g$. In terms of the Majoranas, the product $Z_x Z_y$ is (assuming $y > x$)
\begin{equation}
	Z_x Z_y = i^{(y-x)} \gamma_{2x} \ldots \gamma_{2y-1} .
\end{equation}
To evolve this product to imaginary time $\beta/2$, we insert factors of $e^{\beta H / 2} e^{-\beta H/2}$ between each fermion, and use the identity
\begin{equation}
	e^{\tau H} \gamma_i e^{-\tau H} = \sum_j [e^{-i A \tau}]_{ij} \gamma_j ,
\end{equation}
which follows easily from the Campbell-Baker-Haussdorf formula. Since each $\gamma_i(\tau) \equiv e^{\tau H} \gamma_i e^{-\tau H}$ is simply a linear combination of Majoranas, we can in principle compute the expectation value of any product of Majoranas using Wick's theorem \cite{molinari2023noteswickstheoremmanybody}:
\begin{equation}
	\begin{split}
		\expval{\gamma_{i_1}(\tau_1) \ldots \gamma_{i_{2n}}(\tau_{2n})}_{\beta} &\equiv \frac{1}{\tr e^{-\beta H}} \tr \qty[ \gamma_{i_1}(\tau_1) \ldots \gamma_{i_{2n}}(\tau_{2n}) e^{-\beta H} ] \\
		&= \sum_{\text{Pairings }P} (-1)^{\abs{P}} \expval{\gamma_{i_{P(1)}}(\tau_{P(1)}) \gamma_{i_{P(2)}}(\tau_{P(2)})}_{\beta} \ldots \expval{\gamma_{i_{P(2n-1)}}(\tau_{P(2n-1)}) \gamma_{i_{P(2n)}}(\tau_{P(2n)})}_{\beta} .
	\end{split}
\end{equation}
Importantly, the expectation values required to compute $R_1$ are all very high-order expectation values, with a large number of contractions. We can nevertheless compute such expectation values by converting the above sum over pairings into the Pfaffian of a matrix. Explicitly, define an antisymmetric matrix $\mathcal{G}_{ab}$ with $a, b = 1, \ldots , 2n$ via $\mathcal{G}_{ab} = -\mathcal{G}_{ba} = \expval{\gamma_{i_a}(\tau_a) \gamma_{i_b}(\tau_b)}_{\beta}$. Then Wick's theorem as written above can be reorganized as follows:
\begin{equation}
	\begin{split}
		\expval{\gamma_{i_1}(\tau_1) \ldots \gamma_{i_{2n}}(\tau_{2n})}_{\beta} &= \sum_{\text{Pairings }P} (-1)^{|P|} \mathcal{G}_{P(1), P(2)} \ldots \mathcal{G}_{P(2n-1), P(2n)} \\
		&= \frac{1}{2^n n!} \sum_{\sigma \in S_{2n}} (-1)^{\abs{\sigma}} \mathcal{G}_{\sigma(1),\sigma(2)} \ldots \mathcal{G}_{\sigma(2n - 1), \sigma(2n)} \\
		&= \text{Pf}(\mathcal{G}) .
	\end{split}
\end{equation}
The Pfaffian can then be computed efficiently using the algorithm of Ref.~\cite{wimmer2012algorithm}.

Figure \ref{fig:ising_plots} provides our numerical results for the R\'enyi-1 correlator in the parity-even Gibbs state \eqref{eq:rho_beta_evenparity}. For simplicity we work in a system of size $L = 128$ and fix $\abs{x-y} = L/2$; larger system sizes and more detailed numerical studies are easily accessible. Each curve corresponds to a different temperature, with the lowest temperature (darkest black line) given by $T = 0.15J$, while the highest temperature (brightest red line) corresponds to $T = 2.95J$. We see that $R_1(L/2)$ is indeed nonzero for all $T > 0$. Its exponential decay with $g$ at low temperatures is as expected from the analytical solution for $J \to 0$, with a decay constant proportional to the inverse temperature $\beta$. Thus, our numerics support the conclusion that the one-dimensional TFIM exhibits SWSSB at all nonzero temperatures, for all values of $g$.

\end{document}